\documentclass[12pt]{JHEP3}
\usepackage{epsfig}

\def\spose#1{\hbox to 0pt{#1\hss}}
\def\ltapprox{\mathrel{\spose{\lower 3pt\hbox{$\mathchar"218$}}
 \raise 2.0pt\hbox{$\mathchar"13C$}}}
\def\gtapprox{\mathrel{\spose{\lower 3pt\hbox{$\mathchar"218$}}
 \raise 2.0pt\hbox{$\mathchar"13E$}}}

\title{Spectrum of confining strings in SU($N$) gauge theories
}

\author{Luigi Del Debbio \\ 
	Dipartimento di Fisica dell'Universit\`a
	di Pisa and I.N.F.N., I-56127 Pisa, Italy \\ 
	E-mail: \email{ldd@df.unipi.it} 
} 
\author{Haralambos Panagopoulos \\
	Department of Physics, University of Cyprus, 
	Nicosia CY-1678, Cyprus \\ 
	E-mail: \email{haris@ucy.ac.cy} 
} 
\author{Paolo Rossi \\
	Dipartimento di Fisica dell'Universit\`a 
	di Pisa and I.N.F.N., I-56127 Pisa, Italy \\ 
	E-mail: \email{rossi@df.unipi.it} 
} 
\author{Ettore Vicari \\ 
	Dipartimento di Fisica dell'Universit\`a 
	di Pisa and I.N.F.N., I-56127 Pisa, Italy \\ 
	E-mail: \email{vicari@df.unipi.it} 
}

\abstract{We study the spectrum of the confining strings in
four-dimensional SU($N$) gauge theories.  We compute, for the SU(4)
and SU(6) gauge theories formulated on a lattice, the string tensions
$\sigma_k$ related to sources with $Z_N$ charge $k$, using Monte Carlo
simulations.  Our results are consistent with the sine formula
$\sigma_k/\sigma = \sin k \frac{\pi}{N} / \sin \frac{\pi}{N}$ for the
ratio between $\sigma_k$ and the standard string tension $\sigma$.
For the SU(4) and SU(6) cases the accuracy is approximately 1\% and
2\%, respectively.  The sine formula is known to emerge in various
realizations of supersymmetric SU($N$) gauge theories.  On the other
hand, our results show deviations from Casimir scaling.  We also
discuss an analogous behavior exhibited by two-dimensional SU($N$)
$\times$ SU($N$) chiral models.  }


\begin{document}

\section{Introduction}
\label{sec-intro}

Quantum Chromodynamics
is a nonabelian gauge theory based on the gauge group SU(3). The
mechanisms underlying many of its
fundamental properties, such as confinement, chiral symmetry, topological effects
and the axial anomaly, are under active investigation; they are being studied
by different approaches, including numerical simulations of the theory
formulated on the lattice, 
several models of the vacuum,
as well as some recent proposals derived from
M-theory and AdS/CFT.  
Many features of QCD can be better understood by extending 
the study to SU($N$) gauge theories with $N$ larger than three.
In particular the large-$N$ limit,
which is obtained keeping $g^2 N$ fixed ($g$ is the gauge
coupling) \cite{Hooft-74}, 
is of considerable interest from a phenomenological
point of view, and is one of our best nonperturbative means of
investigating QCD 
(see e.g. the reviews \cite{Das-87,Polyakov-88,PCV-98,Manohar-98}
and references therein).
Indeed, this limit is expected to preserve
qualitatively most nonperturbative features of QCD.

Four-dimensional non-Abelian gauge theories exhibit confinement, i.e. 
static sources in the fundamental representation develop
a linear potential  characterized by a string tension $\sigma$.
As pointed out in many studies, it is important to investigate
the behavior of the system in the presence of static sources in
representations higher than the fundamental one. 
This may provide useful hints on the mechanism
responsible for confinement, helping to identify the most appropriate
models of the QCD vacuum and to select among the various confinement
hypotheses.

SU($N$) gauge theories confine by means of chromoelectric flux tubes carrying charge in the
center $Z_N$ of the gauge group. A chromoelectric source of charge $k$
with respect to  $Z_N$ is confined by a $k$-string with string tension
$\sigma_k$ ($\sigma_1\equiv \sigma$ is the string tension related to 
the fundamental representation). 
If $\sigma_k < k\, \sigma$, then a string with
charge $k$ is stable against decay to $k$ strings of unit charge.
Charge conjugation implies 
$\sigma_k=\sigma_{N-k}$. Therefore $SU(3)$ has
only one independent string tension determining the large distance
behavior of the potential for $k\ne 0$.
One must consider larger values of $N$
to search for distinct $k$-strings.
The spectrum of the $k$-string tensions is then 
determined by the ratios
\begin{equation}
R(k,N) \equiv  {\sigma_k\over \sigma}.
\end{equation}

It has been noted \cite{Strassler-98} that
stable $k$-strings are related to the totally antisymmetric 
representations of rank $k$, and that 
in various realizations of supersymmetric SU($N$) gauge theories
$R(k,N)$ satisfies the sine formula $R(k,N)=S(k,N)$ where
\begin{equation}
S(k,N) \equiv { \sin (k\pi/N) \over  \sin (\pi /N)}\,.
\label{sinf}
\end{equation}
$R(k,N)$ has been computed for the ${\cal N} = 2$ supersymmetric
SU($N$) gauge theory softly broken to ${\cal N} = 1$
\cite{DS-95,HSZ-98}, obtaining Eq. (\ref{sinf}).  The same result has
been found also in the context of M-theory, and extended to the case
of large breaking of the ${\cal N}=2$ supersymmetric theory
\cite{HSZ-98}. The same formula has been recently
rederived~\cite{HK-01} using a different setup, i.e. gauge/string
duality, suggesting that in the ${\cal N}=1$ supersymmetric gauge
theories the sine formula may be quite robust.  The interesting
question is whether the sine formula holds also in nonsupersymmetric
SU($N$) gauge theories.  The M-theory approach to nonsupersymmetric
QCD, although it is still at a rather speculative stage, suggests that
this may be so \cite{Witten-97,HSZ-98}. However, as discussed in
Refs.~\cite{HSZ-98,Strassler-98}, corrections from various sources
cannot be excluded, so that this prediction cannot be considered
robust.

As pointed out in Ref.~\cite{Strassler-98}, it is interesting
to compare the $k$-string tension ratios 
in different theories. The idea is that such ratios may reveal
a universal behavior within  a large class of models characterized by
SU($N$) symmetry, such as SU($N$) gauge theories and their
supersymmetric extensions. 
Therefore, according to this universality hypothesis, 
the $k$-string tension ratios in 
four-dimensional SU($N$) gauge theories should be given by the sine
formula (\ref{sinf}).
This notion of universality for the behavior of the
$k$-string tensions might complement the one conjectured for 
the type of effective string theory describing 
confining strings in gauge theories \cite{LSW-80,CFGHP-97}.

Another interesting and suggestive hypothesis is that the $k$-string
tension ratio satisfies 
the so-called Casimir scaling law \cite{AOP-84}, i.e. 
$R(k,N)=C(k,N)$ where
\begin{equation}
C(k,N) \equiv {k(N-k)\over N-1}
\label{casf}
\end{equation}
is the ratio between the values of the quadratic Casimir
operators in the rank-$k$ antisymmetric and in the 
fundamental representations.
Casimir scaling is satisfied on the one hand  
by the small-distance behavior of the potential between two static charges in different
representations, as shown by perturbation theory up to two
loops~\cite{pert-pot},  and on the other hand by 
the strong-coupling limit of the lattice Hamiltonian formulation
of SU($N$) gauge theories 
\cite{KS-75,Kogut-83,Creutz-85}.
Interest in Casimir scaling was recently
revived~\cite{SS-00,LT-01,KITS-01,FGO-98}; it has been
triggered by  numerical studies of SU(3) lattice gauge
theory \cite{Bali-00,Deldar-00}, which indicate that Monte Carlo 
data for the potential between charges in different
representations are consistent with Casimir scaling
up to a relatively large distance,
$r \approx 1 {\rm fm}$. 

The Casimir scaling law holds exactly in two-dimensional QCD. 
In higher dimensions no strong arguments exist in favor of
a mechanism preserving Casimir scaling 
from small
distance (essentially perturbative, characterized by a 
Coulombic potential) to large distance
(characterized by a string tension for sources carrying $Z_N$
charge); nor across the roughening
transition, from strong to weak coupling. 
We will show explicitly that Casimir scaling
does not survive the next-to-leading order calculation of
the ratios $R(k,N)$ in the strong-coupling lattice Hamiltonian
approach.

It is worth mentioning another simple model for
the spectrum of confining strings:
if the interaction between fundamental flux tubes were
so weak that no bound string states existed, 
then the spectrum would be given by 
\begin{equation}
F(k,N)\equiv {\rm Min}[ k, N-k].
\label{freest} 
\end{equation}

Note that all the above hypotheses considered have the same large-$N$
limit, i.e.
\begin{equation}
S(k,\infty) = C(k,\infty) = F(k,\infty) = k,
\end{equation}
which is the expected result, since no bound states should exist
for $N=\infty$.
Note also that
\begin{equation}
S(k,N) = k + O\left( {1/ N^2}\right).
\end{equation}
In this respect the sine formula is peculiar
because there are no a priori reasons for the large-$N$ expansion of the
$k$-string tension ratio
to be even in $1/N$. 

Of course, it is possible, and even likely given  the current
state of the theoretical knowledge, that none of the above hypotheses
is correct. Nevertheless, we believe that a study able to discard
some of them and determining the size of the corresponding corrections
would be already important for the understanding of
confinement in SU($N$) gauge theories.

The issue of the $k$-strings can be investigated numerically using the
lattice formulation of SU($N$) gauge theories.  Recent numerical
results for $R(2,N)$, obtained for $N=4,5$~\cite{LT-01,WO-01}, show
that $R(2,N)<2\,$; thus, $\sigma_2 < 2\sigma$, indicating that flux
tubes attract each other, and definitely discarding the hypothesis
(\ref{freest}) of free strings.  The available estimates of $R(2,N)$
are substantially consistent with both the sine and Casimir formulas,
thus they do not allow one to exclude any of the two hypotheses.  This
is also due to the fact that the two predictions for $k=2$ are
numerically close, so that high accuracy is necessary to distinguish
them.  In particular, the most precise result for the ratio $R(2,4)$,
reported in Ref.\cite{LT-01}, lies between the predictions from the
sine formula and the Casimir scaling, and is consistent with both
within two error bars.

The aim of this paper is to further investigate this issue.  We
present results from Monte Carlo simulations of the SU($N$) 
lattice gauge theories with $N{=}4,6$  using the Wilson formulation.  
For $N=4$ two independent $k$-strings are expected, including the
fundamental one. For $N=6$ there are three.
We anticipate here our final results for the  $k$-string tension
ratios:
\begin{eqnarray}
R(2,4) &=& 1.403 \pm 0.015,\\
R(2,6) &=& 1.72 \pm 0.03 ,\\
R(3,6) &=& 1.99 \pm 0.07 .
\end{eqnarray}
Moreover, we found  no evidence for stable string states
 associated with the symmetric rank-2 representation,
in accordance with general arguments.

Figure~\ref{summary} summarizes our results, comparing our MC results
with the above-mentioned hypotheses of spectrum.
We claim that 
SU(4) and SU(6) results show substantial agreement with the sine
formula, and therefore with the
universality conjecture  for the spectrum of the confining strings 
in asymptotically free theories with SU($N$) symmetry.
The sine formula (\ref{sinf}) predicts $S(2,4)=\sqrt{2}=1.414...$, $S(2,6)
= 1.732...$, and $S(3,6)=2$. 
Moreover, the results show deviations from a strict  Casimir scaling,
whose 
predictions are $C(2,4)=4/3$, $C(2,6)=8/5$ and $C(3,6)=9/5$.

Considering our results all together, we can state that the sine
formula is consistent within an accuracy of approximately 1\%.  This
fact should be relevant for the recent debate on confinement models,
such as those discussed in Refs.
\cite{SS-00,LT-01,KITS-01,FGO-98,DFGO-97,HV-98,ScSu-00,Yung-00,EFMG-00,KB-01,Deldar-01,KS-01}.
Of course, our numerical results cannot prove that the sine formula
holds exactly, but they place a very stringent bound on the size of
the possible corrections.  At the same time, our results appear rather
conclusive on the existence of deviations from the Casimir scaling.
Casimir scaling may still be considered as a reasonable approximation,
since the largest deviation is about 10\% for $R(3,6)$.
\FIGURE[ht]{
\epsfig{file=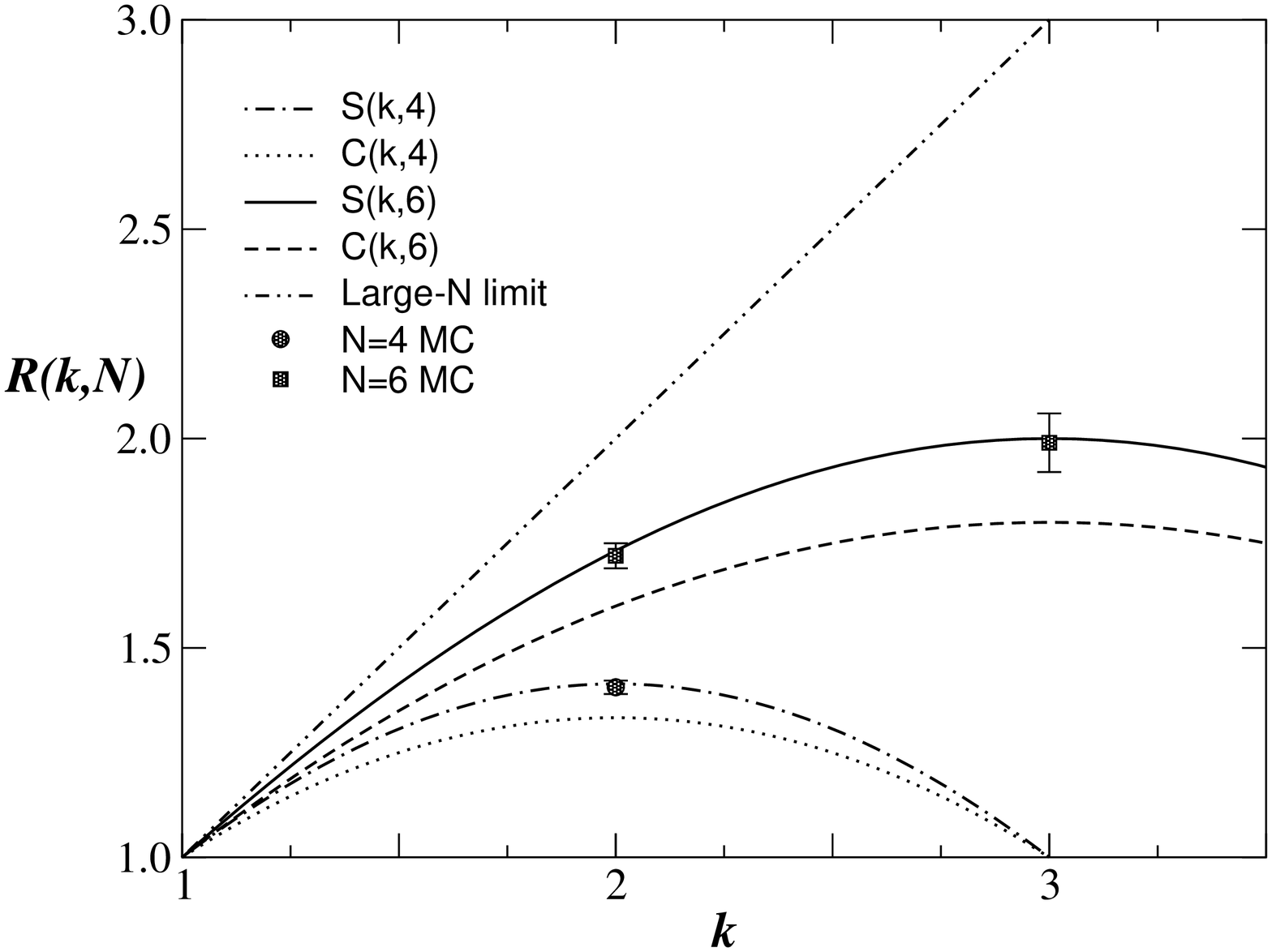, width=12truecm}
\caption{Comparison of the various hypotheses for the $k$-string
ratios with the Monte Carlo results.}
\label{summary}
}

Finally, it is interesting to note that the sine formula (\ref{sinf})
also emerges in the context of the two-dimensional SU($N$)$\times$
SU($N$) chiral models (see e.g. Ref.~\cite{Polyakov-88} as a general
reference).  $d$-dimensional chiral models and $2d$-dimensional
lattice gauge theories present interesting analogies. In particular,
the relation is exact for $d=1$, and one can prove that Casimir
scaling holds for the masses of the bound states.  In analogy with
four-dimensional SU($N$) gauge theories, in two-dimensional
SU($N$)$\times$ SU($N$) chiral models the Casimir scaling law holds
for the small-distance behavior of the correlation functions related
to different representations.  Moreover, it also holds for the
strong-coupling limit of the corresponding lattice Hamiltonians, but
it is not satisfied at next-to-leading order for a generic choice of
the lattice Hamiltonian, such as the one of Ref.~\cite{SK-81}. On the
other hand, the exact S-matrix, derived using essentially the Bethe
Ansatz \cite{AAL-84,Wiegmann-84}, shows that bound states exist only
for the rank-$k$ antisymmetric representations, and the ratio of their
masses satisfies the sine formula.

The paper is organized as follows. In Sec.~\ref{sec4} we present the
results of the Monte Carlo simulations of the SU(4) and SU(6) lattice
gauge theories.  The rest of the paper presents analytical results
that help in providing a more detailed picture of the characteristic
features of the potential between static charges in higher-rank
representations. In Sec.~\ref{sec3} we describe the computation of the
strong-coupling expansion in the lattice Hamiltonian approach of
SU($N$) gauge theories, to the first nontrivial next-to-leading order.
We show explicitly that Casimir scaling is violated by the corrections
to the leading order.  In Sec.~\ref{sec2} we discuss the analogies
between chiral models and lattice gauge theories.  The appendices are
dedicated to a number of issues related to the Monte Carlo simulations
of SU($N$) gauge theory at large $N$, such as the matching of the
lattice couplings in the large-$N$ limit, the bulk transition observed
at finite bare coupling for $N$ sufficiently large, and the severe
form of critical slowing down which characterizes the Monte Carlo
dynamics of the topological quantities such as the topological charge,
and which appears to follow an exponential law rather than a power
law.

Short reports containing essentially our Monte Carlo results for the SU(6) gauge theory,
and some preliminary results for SU(4), have already
appeared in Refs.~\cite{DPRV-01,DPRV-01-2}.

\section{$\lowercase{k}$-strings in four-dimensional SU(4) and SU(6) gauge theories}
\label{sec4}

In order to investigate the behavior of the $k$-string tensions in
gauge theories, we performed numerical Monte Carlo simulations of the
four-dimensional lattice SU(4) and SU(6) gauge theories using their Wilson
formulation  
\begin{equation}
S_{\rm gauge} = - N\beta \sum_{x,\mu>\nu} {\rm Tr} \left[
U_\mu(x) U_\nu(x+\mu) U_\mu^\dagger(x+\nu) U_\nu^\dagger(x) 
+ {\rm h.c.}\right].
\label{wilsonac}
\end{equation}

In our simulations we employed the Cabibbo-Marinari algorithm
\cite{CM-82} to upgrade SU($N$) matrices by updating their SU(2)
subgroups (we selected 6 and 15 subgroups respectively
for the SU(4) and SU(6) cases). This was done by alternating
microcanonical over-relaxation and heat bath steps, typically in a 4:1
ratio.  In the following we consider a sweep as the upgrading
of all links of the lattice independently of the algorithm;
thus a over-relaxation and heat-bath cycle takes 5 sweeps.
In Tables \ref{tab:su4data} and \ref{tab:su6data} we present some information on our
Monte Carlo runs for the SU(4) and SU(6) cases respectively. 
We provide the coupling value 
\begin{equation}
\gamma\equiv {\beta \over 2N^2}
\label{gammdef}
\end{equation}
(this rescaled coupling is more natural for large $N$, due to the fact
that the large-$N$ limit of the lattice theory is obtained  keeping
$\gamma$ fixed),   the size of the lattices
and the number of measurements, reported as the ratio between the
number of sweeps $N_{\rm sw}$ and the interval between two
measurements $N_{\rm m}$. 
Moreover we report the values of the
mean field \cite{improved-couplings,LM-93} 
and cactus \cite{PV-98} improved couplings, 
$\gamma_{\rm mf}$ and $\gamma_{\rm cactus}$ respectively, and 
the value of the lattice spacing in units of the square root of the
string tension. As discussed in App.~\ref{appa},  
these quantities are useful to compare lattice results
at different values of $N$.
In total, the whole study took about 10 years of CPU 
on a Pentium III 1Ghz cluster.
\TABLE[ht]{
\caption{Data sets available for SU(4).}
\label{tab:su4data}
\begin{tabular}{cccccll}
\multicolumn{1}{c}{$\gamma$}&
\multicolumn{1}{c}{$\gamma_{\rm cactus}$}&
\multicolumn{1}{c}{$\gamma_{\rm mf}$}&
\multicolumn{1}{c}{lattice}&
\multicolumn{1}{c}{$N_{\rm sw}/N_{\rm m}$}&
\multicolumn{1}{c}{$a \sqrt{\sigma}$}&
\multicolumn{1}{c}{$L\sqrt{\sigma}$}\\
\hline
0.335  & 0.24196  & 0.1862 & $ 12^3\times 24$ & 2047k/10  & 0.2959(14) & 3.55 \\ 
0.337  & & & $ 16^3\times 32$ & 2290k/20  & 0.2699(23)  & 4.32 \\
0.338  & 0.24523  & 0.1906 & $ 12^3\times 24$ & 3858k/20  & 0.2642(7)  & 3.17 \\
0.341  & 0.24850  & 0.1947 & $ 12^3\times 24$ & 4308k/20  & 0.2368(6) & 2.84 \\ 
0.344  & 0.25176  & 0.1987 & $ 12^3\times 24$ & 2018k/20  & 0.2122(8) & 2.55 \\ 
       & & & $ 16^3\times 32$ & 3615k/20  & 0.2160(8)& 3.46 \\ 
0.347  & & & $ 16^3\times 32$ & 4674k/20  & 0.1981(5) & 3.17 \\ 
\end{tabular}
}
\TABLE[ht]{
\caption{Data sets available for SU(6).}
\label{tab:su6data}
\begin{tabular}{cccccll}
\multicolumn{1}{c}{$\gamma$}&
\multicolumn{1}{c}{$\gamma_{\rm cactus}$}&
\multicolumn{1}{c}{$\gamma_{\rm mf}$}&
\multicolumn{1}{c}{lattice}&
\multicolumn{1}{c}{$N_{\rm sw}/N_{\rm m}$}&
\multicolumn{1}{c}{$a \sqrt{\sigma}$}&
\multicolumn{1}{c}{$L\sqrt{\sigma}$}\\
\hline
0.342  & 0.24234 & 0.1843  & $ 8^3\times 16$ & 213k/10  & 0.3151(6) & 2.52 \\ 
       &         &         & $12^3\times 24$ & 520k/20  & 0.3239(8) & 3.89 \\
0.344  & 0.24455 & 0.1875  & $12^3\times 24$ & 727k/20  & 0.2973(5) & 3.57 \\
0.348  & 0.24897 & 0.1935  & $10^3\times 20$ & 592k/20  & 0.2534(6) & 2.53 \\ 
       &         &         & $12^3\times 24$ & 712k/20  & 0.2535(6) & 3.04 \\ 
0.350  & 0.25117 & 0.1963  & $12^3\times 24$ & 442k/20  & 0.2380(6) & 2.86 \\ 
0.354  & 0.25556 & 0.2017  & $12^3\times 24$ & 270k/20  & 0.2103(5) & 2.52 \\ 
\end{tabular}
}

The couplings were chosen to lie in the weak-coupling region.
This is important because the Wilson lattice formulations
of SU($N$) gauge theories undergo a first order phase transition for $N$
sufficiently large, as argued using various approaches, such as Monte Carlo
simulations \cite{Creutz-81,CrMo-82,others-PT,IIY-95},
mean field calculations \cite{DZ-83,ID-89}, 
reduced models \cite{Campostrini-99}.
We found evidence for a first order phase transition in the SU(6) case,
for  $\gamma_c=0.3389(4)$.  
This issue is discussed in App. \ref{appb}, where some additional Monte Carlo results
are presented.  Therefore, in our simulations
we considered $\gamma$ values larger than  $\gamma_c$. Moreover, in order to avoid
getting trapped in unphysical metastable states, we always used cold
configurations as the starting point of our simulations. 
On the other hand, in the SU(4) case 
the MC data  of the specific heat 
do not show any evidence for a bulk transition,
contrary to some expectations  coming from  mean field calculations
\cite{ID-89} and earlier Monte Carlo simulations 
\cite{Creutz-81,others-PT}. The crossover between the strong- and
weak-coupling region is characterized
by a pronounced peak of the specific heat at $\gamma\simeq 0.325$
(corresponding to $\beta\simeq 10.4$), similarly to the SU(3) case
where the absence of a bulk transition is well established.

We used asymmetric lattices ($L^3\times T$) with a larger time size,
along which the wall-wall correlations of Polyakov loops were measured.  
For some values of
$\gamma$ we performed simulations for two lattice sizes in order to
check for finite size effects.  The lattice sizes $L$ were chosen so
that $L\sqrt{\sigma}\gtapprox 2.5$, and for most of them
$L\sqrt{\sigma}\simeq 3$. 
This requirement ensures that finite size effects on $k$-string ratios
are negligible, as can be seen by comparison of the results for different
sizes (see also Refs.~\cite{LT-01-2}). 

In our simulations we measured also the topological charge $Q$, by a
cooling technique, see e.g. Ref.~\cite{cooling}. A severe form of
critical slowing down is observed in this case, which worsens with
increasing $N$.  Estimates of the autocorrelation time $\tau_Q$ for
$Q$, obtained from a blocking analysis of the data, turns out to be
consistent with an exponential increase: $\tau_Q\propto \exp (c
\xi_\sigma)$ where $\xi_\sigma\equiv \sigma^{-1/2}$, with $c\approx
1.7$ for SU(4) and $c\approx 2.5$ for SU(6).  As a consequence, the
run for the largest value of $\gamma$ we considered for SU(6),
i.e. $\gamma=0.354$, did not correctly sample $Q$, presumably because
it was not sufficiently long ($\approx$ 300k sweeps).  This dramatic
effect was not observed in the correlators used to determine the
$k$-string tensions, suggesting an approximate decoupling between the
topological and nontopological modes.  Indeed a blocking analysis did
not show significant autocorrelations in measurements taken every
10-20 sweeps.  In App.~\ref{appc} this issue is discussed in more
detail.  The critical slowing down shown by the topological quantities
represents a severe limitation for numerical studies of the
topological properties at large values of $N$ using standard Monte
Carlo algorithms.  The results related to the topological properties
will be reported elsewhere.

The $k$-string tensions are extracted from the large-time behavior of
correlators of strings in the antisymmetric representations, closed
through the periodic boundary conditions (see e.g.
Refs.\cite{DSST-85,LT-01}):
\begin{equation}
C_r(t) = \sum_{x_1,x_2} \langle \chi_r [ P(0;0) ]\; 
\chi_r [ P(x_1,x_2;t) ] \rangle,
\end{equation}
where 
\begin{equation}
P(x_1,x_2;t) = \Pi_{x_3} U_3(x_1,x_2,x_3;t).
\end{equation}
$U({\bf x}; t)$ are the usual link variables, and $\chi_r$ is the
character of the representation $r$. In particular for the fundamental
representation: 
\begin{equation}
\chi_{f}[ P ] = {\rm Tr}\, P,
\end{equation}
for the antisymmetric representation of rank $k=2$
\begin{equation}
\chi_{k=2}[ P ] = {\rm Tr}\, P^2 - \left( {\rm Tr} \,P\right)^2,
\end{equation}
and for the antisymmetric representation of rank $k=3$ 
\begin{equation}
\chi_{k=3}[ P ] = 2 {\rm Tr}\, P^3 - 3 {\rm Tr}\, P^2 \; {\rm Tr}\,P +
\left( {\rm Tr} \,P\right)^3.
\end{equation} 

These correlators decay exponentially as $\exp (- m_k t)$ where $m_k$ is the
mass of the lightest state in the corresponding representation.
For a $k$-loop of size $L$, the $k$-string tension is obtained
using the relation \cite{DSST-85}
\begin{equation}
m_k = \sigma_k L - {\pi\over 3 L}.
\label{mks}
\end{equation}
The last term in Eq.~(\ref{mks}) is conjectured to be a universal
correction, and it is related to the universal critical behavior of
the flux excitations described by a free bosonic string \cite{LSW-80}.
Numerical results for the three- and four-dimensional $SU(N)$ gauge
theories with various values of $N$, see
e.g. Refs. \cite{LT-01,LT-01-2,Teper-98}, and for the
three-dimensional $Z_2$ gauge theory \cite{CFGHP-97} support a
universal description of the flux excitations in terms of a free
bosonic string.  So it is reasonable to assume that this picture, and
therefore the relation (\ref{mks}), be valid every time that a stable
string state propagates, regardless of the gauge group or the
representation considered. The comparison of the results obtained for
different lattice sizes will give further support to this assumption.
Note that Eq.~(\ref{mks}) is expected to hold for sufficiently large
values of $L$.  Ref. \cite{LT-01-2} argues that a lattice size $L$
satisfying $L\sqrt{\sigma}\gtapprox 3$ should be sufficient to observe
a behavior according to Eq.~(\ref{mks}) for the loop mass. (Although
Ref.~\cite{LT-01-2} considered SU($N$) gauge theories with $N=4,5$,
the case $N=6$ should be equivalent in this respect.)

In order to improve the efficiency of the measurements we used
smearing and blocking procedures (see e.g. Refs.~\cite{smearing}) to
construct new operators with a better overlap with the lightest string
state. The smearing procedure replaces every spatial link on
the lattice according to:
\begin{equation}
U_k(x) \mapsto {\cal P} \left\{ U_k(x) + \alpha_s \sum_{\pm(j\neq k)} U_j(x)
U_k(x+\hat j) U^\dagger_j(x+\hat k) \right\},
\end{equation}
where ${\cal P}$ indicates the projection onto $SU(N)$ and the sum
only runs on spatial directions. The blocking procedure replaces the
spatial links with super-links ${\cal U}_k(x)$ defined on a lattice
with lattice spacing $2a$ (except for the time direction) according to
\begin{equation}
{\cal U}_k(x) = 
{\cal P} \left\{ U_k(x) U_k(x+\hat k) + 
\alpha_f \sum_{\pm(j\neq k)} U_j(x)
U_k(x+\hat j) U_k(x+\hat j+\hat k) U^\dagger_j(x+2\hat k) \right\}.
\end{equation}
The blocking procedure can then be iterated $n$ times to produce
super-links of length $2^n a$. The coefficients $\alpha_s$ and
$\alpha_f$ can be adjusted to optimise the efficiency of the
procedure. We constructed new super-links using
$\alpha_s=\alpha_f=0.5$, three-smearing, and a few blocking steps,
according to the value of $L$, i.e. one for $L=10$, two
for $L=8,12$ and three for $L=16$. These
super-links were used to compute improved Polyakov lines. 
\FIGURE[ht]{
\epsfig{file=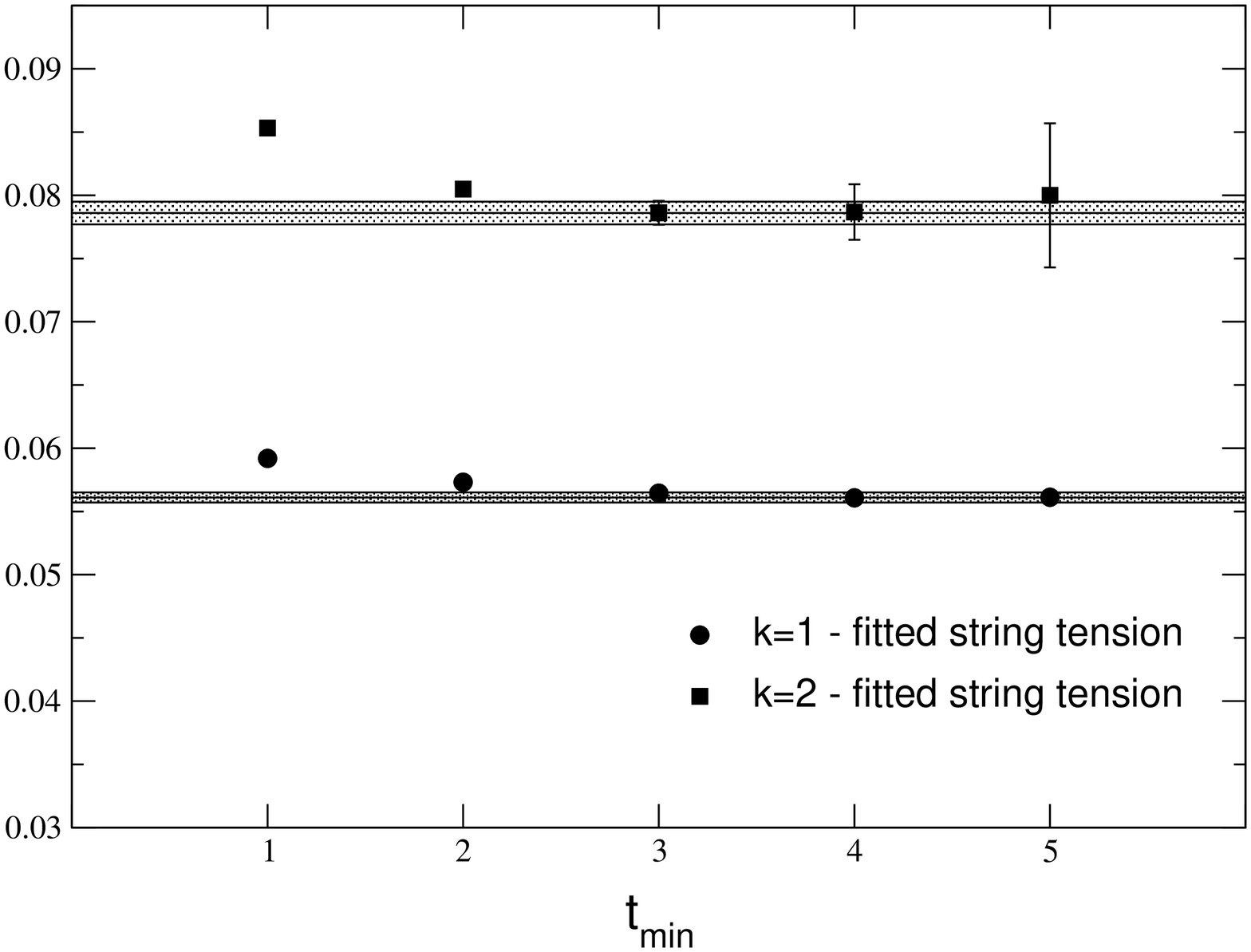, width=12truecm}
\caption{The $k$-string tensions determined by the fits to the
corresponding wall-wall correlations, as functions of the lower bound
of the fitrange, $t_{\rm min}$, for SU(4) and $\gamma=0.341$. We also
show our final estimates, which are represented by the the continuous
lines.}
\label{fitresults}
}

\FIGURE[ht]{
\epsfig{width=12truecm,angle=0,file=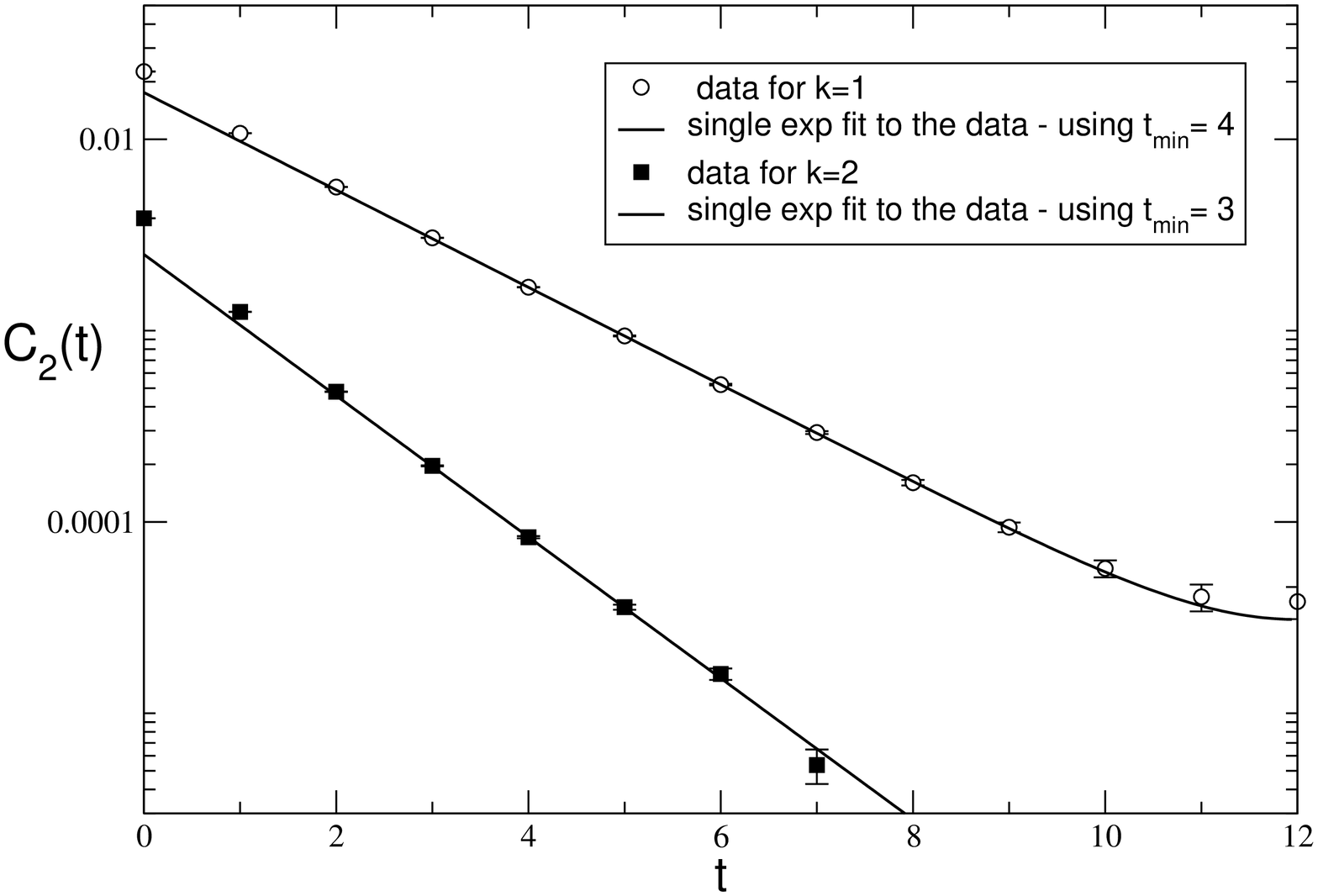}
\caption{Data for the wall-wall correlations whose large-distance
exponential behaviors determine the $k$-strings, for SU(4) and
$\gamma=0.341$.}
\label{fit}
}

We used a standard blocking analysis to check for possible
autocorrelations in the wall-wall correlators used to determine the
masses.  The masses $m_k$ were obtained from fitting the time
behaviour of the folded 2-pt correlators:
\begin{equation}
F_{\rm folded}(t) \equiv \frac12 \left(F_k(t) + F_k(N_t-t)\right)
 = A (e^{-m_k t} + e^{-m_k(N_t-t)}).
\label{fold}
\end{equation}
The statistical error on the fitted parameters was computed using a
bootstrap procedure. 

The choice of the fit range $[t_{\rm min}, t_{\rm max}]$ is a delicate
issue.  In a lattice computation of masses the choice of the fit
range is a source of systematic error that is very difficult to
control.  On the one hand, the data at early times are still
contaminated by heavier states. On the other hand, the masses are so
large in lattice units that the relative error of data increases
rapidly with the distance.  A large value of $t_{\rm min}$ reduces the
systematic error due to contamination of heavier states, but leads to
an increase of the statistical error. A satisfactory compromise is
reached when the two errors are comparable, or, more cautiously, when
the systematic error is estimated to be negligible with respect to the
statistical one.

The high statistics we collected for the SU(4) lattice gauge theory
allowed us to achieve a good control of the systematic error coming
from the contamination of heavier states.  As an example, we discuss
in some detail the analysis of the data obtained for $\gamma=0.341$,
using approximately $2\times 10^5$ measurements.  In
Fig.~\ref{fitresults} we show the results for $\sigma$ and $\sigma_2$
as obtained varying $t_{\rm min}$, together with the final estimates
that we will quote later.  The value of $t_{\rm max}$ is not critical
in this respect, the data reported in the figure have been obtained
using $t_{\rm max}=8$, but the results are essentially independent of
$t_{\rm max}$.  We observe clearly a dependence on $t_{\rm min}$.  Our
estimates are taken when a plateau is reached, i.e.  for $t_{\rm
min}=4$ in the case of $\sigma$, and $t_{\rm min}=3$ for $\sigma_2$.
This should ensure that the systematic error due to heavier states is
at most comparable with the statistical one.  In Fig.~\ref{fit} we
show the data for the wall-wall correlations corresponding to the
fundamental and $k=2$ antisymmetric representations, and the curves
(\ref{fold}) with our best estimates of the parameters. Thus we
believe that the results and the errors we quote for SU(4) should
account for this systematic error.

For the SU(6) lattice gauge theory we could not afford such a clean
analysis because of the relatively limited statistics.  In this case
fits were typically performed in the range $[2-4]$ for $\gamma \ge
0.348$, and $[1-4]$ for the other values of the coupling.  The
systematic errors were checked by comparing the outcomes of fits over
different time ranges.  In practice, we could not use data for
distances larger than $t=5$, which did not allow us to check the
dependence on the fit range as satisfactorily as in the SU(4) case.
Thus the values of the $k$-string tensions may be still subject to a
systematic error due to the contamination of heavier states.  However,
we note that the ratios $R(k,N)$ turn out to be more stable with
respect to the choice of the fit range (this is already apparent from
Fig.~\ref{fitresults}). We indeed found that the variations of the fit
range yield consistent results within the statistical error.  So the
estimates and the errors that we finally report for the ratios
$R(k,N)$ should be reliable also in the SU(6) case.
\TABLE[htb]{
\caption{$k$-string tensions for SU(4). 
}
\label{tab:results4}
\begin{tabular}{cclll}
\multicolumn{1}{c}{$\gamma$}&
\multicolumn{1}{c}{lattice}&
\multicolumn{1}{c}{$a^2\sigma$}&
\multicolumn{1}{c}{$a^2\sigma_2$}&
\multicolumn{1}{c}{$\sigma_2/\sigma$}\\
\hline
0.335   & $ 12^3\times 24$ & 0.0876(8) & 0.1203(17) & 1.374(20) \\
0.337   & $ 16^3\times 32$ & 0.0728(12)& 0.103(3)   & 1.413(38) \\
0.338   & $ 12^3\times 24$ & 0.0698(4) & 0.0973(12) & 1.395(18) \\ 
0.341   & $ 12^3\times 24$ & 0.0561(3) & 0.0786(10) & 1.402(17) \\ 
0.344   & $ 12^3\times 24$ & 0.0450(3) & 0.0634(7)  & 1.407(15) \\ 
        & $ 16^3\times 32$ & 0.0466(3) & 0.0656(9)  & 1.408(20) \\ 
0.347   & $ 16^3\times 32$ & 0.0392(2) & 0.0549(4)  & 1.400(10) \\ 
\end{tabular}
}
\TABLE[htb]{
\caption{$k$-string tensions for SU(6).}
\label{tab:results6}
\begin{tabular}{cclllll}
\multicolumn{1}{c}{$\gamma$}&
\multicolumn{1}{c}{lattice}&
\multicolumn{1}{c}{$a^2\sigma$}&
\multicolumn{1}{c}{$a^2\sigma_2$}&
\multicolumn{1}{c}{$a^2\sigma_3$}&
\multicolumn{1}{c}{$\sigma_2/\sigma$}&
\multicolumn{1}{c}{$\sigma_3/\sigma$}\\
\hline
0.342   & $ 8^3\times 20$ &0.0993(4)  &0.164(1) &0.190(3)  & 1.65(2) & 1.91(3) \\ 
	& $12^3\times 24$ &0.1049(5)  &0.174(3) &0.201(9)  & 1.66(3) & 1.91(9) \\ 
0.344   & $12^3\times 24$ &0.0884(3)  &0.153(2) &0.173(4)  & 1.73(2) & 1.95(5) \\
0.348   & $10^3\times 20$ &0.0642(3)  &0.111(2) &0.134(7)  & 1.73(3) & 2.08(10) \\  
        & $12^3\times 24$ &0.0642(3)  &0.110(3) &0.132(7) & 1.71(4) & 2.06(11) \\  
0.350   & $12^3\times 24$ &0.0567(3) &0.097(2) &0.110(6)  & 1.72(3) & 1.95(9)\\
0.354   & $12^3\times 24$ &0.0442(2) &0.0766(11) &0.090(3)  & 1.73(3) & 2.04(6)  \\  
\end{tabular}
}

\FIGURE[ht]{
\epsfig{width=12truecm,angle=0,file=Fig4b.eps}
\caption{The scaling ratio $R(2,4)$ as a function of $a^2 \sigma$
for $SU(4)$.}
\label{fig:R24}
}

Results for the $k$-string tensions obtained in our simulations are
reported in Tables \ref{tab:results4} and \ref{tab:results6} for SU(4)
and SU(6), respectively.  The relative errors on the ratios $R(2,4)$,
$R(2,6)$ and $R(3,6)$ are essentially determined by the uncertainty on
$\sigma_2$ and $\sigma_3$, since the estimates of $\sigma$ are in
general much more precise.  A bootstrap analysis on the data performed
directly on the ratios $R(k,N)$, once the fit range has been chosen,
usually provides statistical errors that are smaller than the ones we
report. However we do not consider them sufficiently reliable, since
the systematic error due to the contamination of heavier states is
controlled only within the statistical error of the $k$-string
tensions.  The ratios $R(2,4)$, $R(2,6)$ and $R(3,6)$ are plotted in
Figs.~\ref{fig:R24}, \ref{fig:R26}, and \ref{fig:R36} respectively,
versus $a^2\sigma$ to evidentiate possible scaling corrections, which
are expected to be $O(a^2)$ apart from logarithms.  To facilitate the
comparison, in the figures we show the predictions of the sine formula
(\ref{sinf}) and of the Casimir scaling (\ref{casf}).  Before going
into details, it is worthwhile to emphasize some common trends in the
results.  Data confirm that the finite-size effects on the $k$-string
tension ratios are small for $L\sqrt{\sigma} \gtrsim 2.5$. In all
cases the approach to scaling is satisfactory, and the pattern of the
scaling corrections turns out to be similar in SU(4) and SU(6).

\FIGURE[ht]{
\epsfig{file=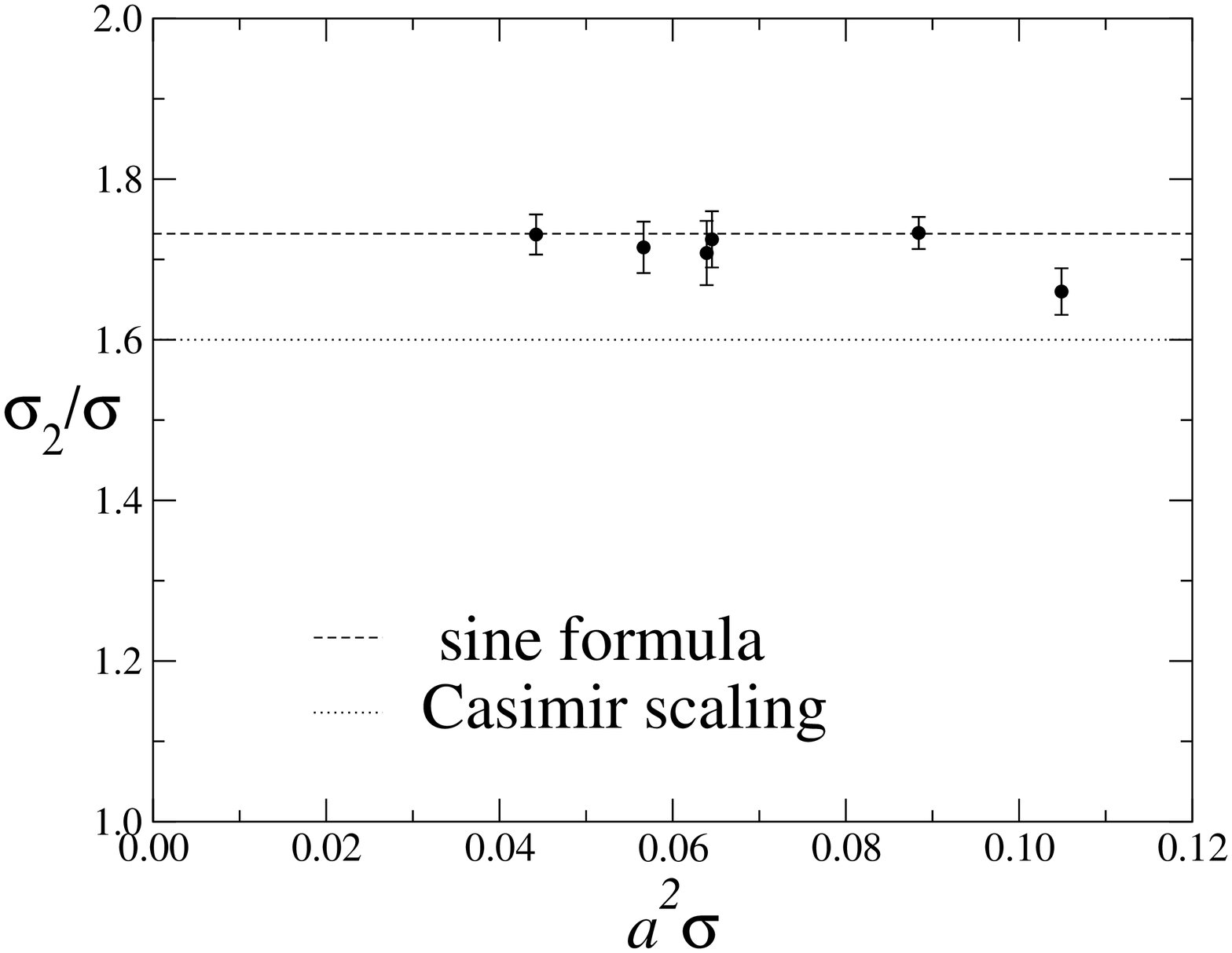, width=12truecm}
\caption{The scaling ratio $R(2,6)$ as a function of $a^2 \sigma$
for $SU(6)$.}
\label{fig:R26}
}

\FIGURE[ht]{
\epsfig{file=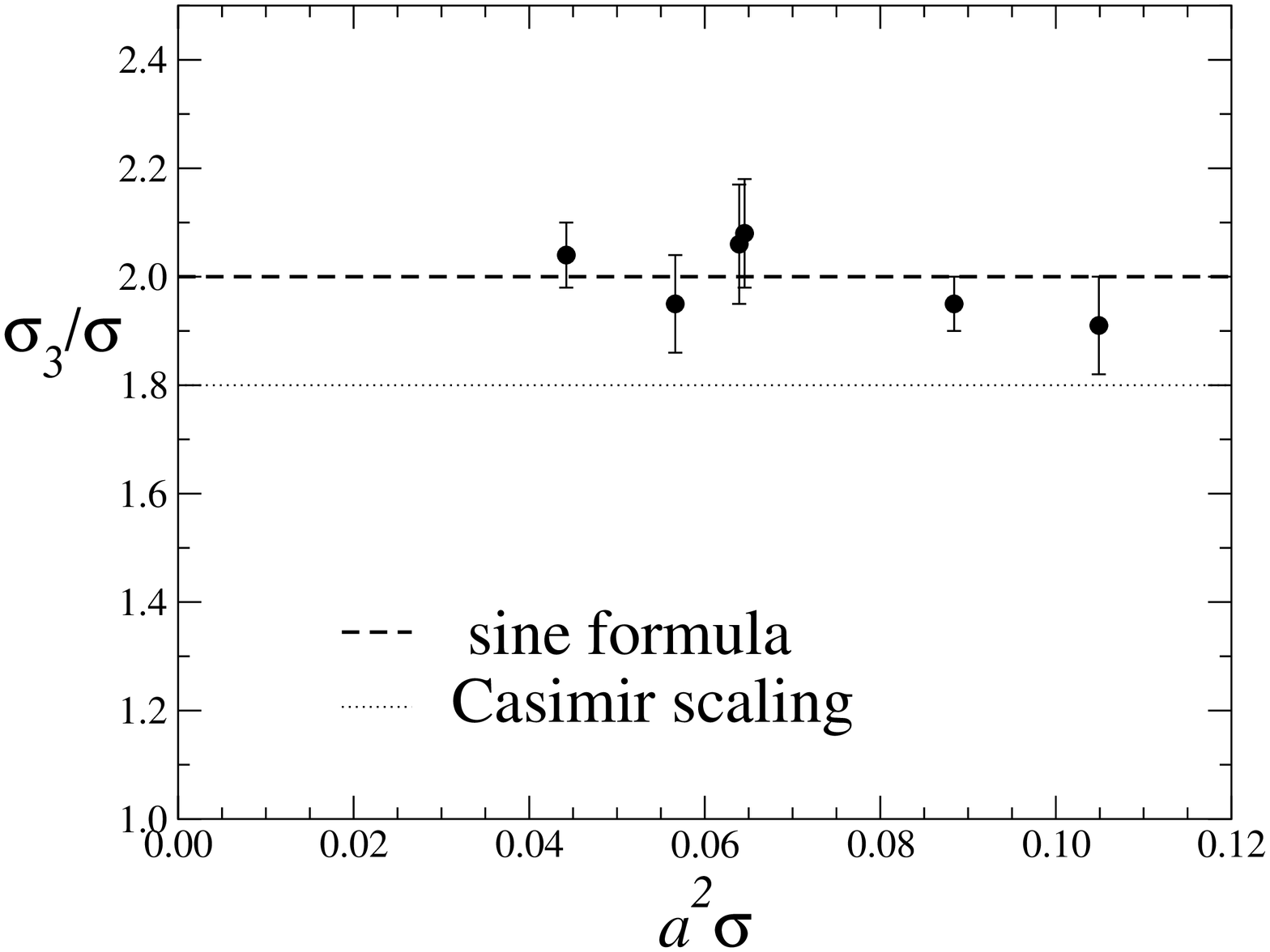, width=12truecm}
\caption{The scaling ratio $R(3,6)$ as a function of $a^2 \sigma$
for $SU(6)$.}
\label{fig:R36}
}

Let us discuss in more detail the results for the SU(4) gauge theory.
Comparing the results obtained at $\gamma=0.344$ for $L=12$ and
$L=16$, we see that when $L\sqrt{\sigma}\gtrsim 2.5$ the size effects
are small, and within the statistical errors of our data for the
$k$-string ratios. The size effects are instead observable within our
statistical errors on $\sigma$ and $\sigma_2$.  The data show a good
scaling behavior.  Only the results for the smallest value of
$\gamma$, i.e. $\gamma=0.335$ apparently show scaling corrections.  We
extract our final estimate for the ratio $\sigma_2/\sigma$ from the
data at the largest values of $\gamma$, i.e.  $\gamma=0.341,0.344,0.347$,
and for their corresponding largest lattices.  Combining them to
obtain the central value and taking the typical error of each single
point as estimate of the error, we propose as final estimate
$R(2,4) =  1.403 \pm 0.015$.
Of course, this estimate assumes that the scaling corrections are
already small and negligible for $a^2\sigma\simeq 0.05$.  The data for
smaller $\gamma$, i.e. $\gamma=0.335,0.338$, are used to check this
hypothesis.  They indicate that the scaling corrections are of the
same size as the error we reported above. We will return to this point
later.  Our result is consistent with the prediction of the sine
formula, $S(2,4) = \sqrt{2}=1.414...$, within an uncertainty of
approximately 1\%.  On the other hand, this result does not support
Casimir scaling, whose prediction in this case is $C(2,4)=4/3$.

The result obtained in Ref.~\cite{LT-01}, i.e. $R(2,4)=1.357(29)$, is
marginally consistent with ours. The comparison improves considering
the results $R(2,4)=1.377(35)$, obtained in Ref.~\cite{LT-01} by
discarding the data for the smallest value of $\beta$,
i.e. $\beta=10.55$ corresponding to $\gamma=0.3297...$.

Let us now consider the results for the SU(6) gauge theory.  Comparing
the results for the string tension ratios for different lattice sizes
at constant $\gamma$ shows little finite size effects.  As in the
SU(4) case, finite size effects are observed on the $k$-string
tensions, but they cancel out in the ratios.  The ratio $R(2,6)$
displays good scaling for $\gamma>0.342$.  Given the good scaling
behaviour, again we do not attempt to fit the dependence of our result
on the lattice spacing $a$. Our final value for $R(2,6)$ is obtained
averaging the results at $\gamma=0.348$ and $\gamma=0.350$, while the
error is given by the typical error of each single point.  Analogously
to the SU(4) case, the data at smaller $\gamma$, and in particular the
one at $\gamma=0.344$, are used to check that the scaling corrections
are small, and at most comparable with the error we report.  Given the
poor sampling of the topological charge in the run at $\gamma=0.354$,
we do not include this value of $\gamma$ in the final estimate of the string
tension ratio. Similar comments apply to the $R(3,6)$ ratio.  We
finally note that the results for $R(k,6)$ at $\gamma=0.354$ are in
agreement with those obtained at smaller values of $\gamma$, for which
$Q$ was sampled correctly, lending further support to the previously
mentioned large decoupling of the modes determining the string
tensions and the topological ones.  As anticipated in the
introduction, our final estimates are $R(2,6) = 1.72 \pm 0.03$ and
$R(3,6) = 1.99 \pm 0.07$.  They are both consistent with the
predictions of the sine formula (\ref{sinf}), which are $S(2,6) =
1.732...$ and $S(3,6)=2$, respectively, within an uncertainty of
approximately 2\% for $k=2$ and 4\% for $k=3$.  On the other hand,
they do not support Casimir scaling; the predictions in this case are
$C(2,6)=8/5$ and $C(3,6)=9/5$.

We have also explored correlators in the symmetric rank-2
representation, whose character is given by
\begin{equation}
\chi_{k=2, {\rm symm}}[ P ] = {\rm Tr}\, P^2 + \left( {\rm Tr} \,P\right)^2,
\end{equation}
finding no evidence for new stable string states for both SU(4) and
SU(6) cases.  The masses extracted in the symmetric channel were
consistent with $m_{\rm symm} \ge 2 m_{\rm f}$, as expected because
the symmetric string should indeed decay into two fundamental
strings. In this case, Eq.~(\ref{mks}) should not apply and we did not
try to extract a string tension.

In conclusion, our results turn out to be consistent with the sine
formula, and show that there are sizeable corrections to the Casimir
scaling prediction.  In order to further check the robustness of our
statement concerning the deviation from the Casimir formula, we have
analyzed our data using also fits which allow explicitly for scaling
corrections.  As suggested in Ref.~\cite{LT-01}, one may fit the data
to the linear behavior $A+B\sigma$.  For this purpose one must
consider all data, including the ones for the smallest values of
$\gamma$, since they are the only ones showing apparent scaling
violations.  In the SU(4) case we find $R(2,4)=1.424(23)$ with
$\chi^2\simeq 1.0$.  In the SU(6) case the results are
$R(2,6)=1.76(4)$ ($\chi^2\simeq 2.6$), $R(3,6)=2.13(11)$
($\chi^2\simeq 1.4$) and $R(2,6)=1.77(6)$ ($\chi^2\simeq 2.7$),
$R(3,6)=2.16(19)$ ($\chi^2\simeq 1.3$) respectively with and without
the point at the largest value of $\gamma$, i.e.  $\gamma=0.354$.
Therefore, if one wants to be more cautious in treating the systematic
error due to scaling corrections, one may take into account the
difference between the linear fits and the nonextrapolated data,
yielding
\begin{eqnarray}
&&R(2,4) = 1.403 \pm 0.015\;^{+0.021}_{-0.000}, \\
&&R(2,6) = 1.72\pm 0.03\;^{+0.05}_{-0.00},\\
&&R(3,6) = 1.99\pm 0.07\;^{+0.17}_{-0.00},
\end{eqnarray}
which cover all the results obtained above.  These conservative
estimates are still not consistent with the Casimir formula. Thus, our
conclusions are fully justified even by this overly cautious analysis.
One should note that corrections to Casimir scaling are to be
expected as discussed in the previous section (see also
Sec.~\ref{sec3}). The implications of these results for models of the
Yang-Mills vacuum deserve further investigation.

One may write down a general expression for $R(k,N)$ taking 
into account the following constraints:
$R(1,N)=1$ by definition, $R(k,N)=R(N-k,N)$
by charge conjugation, and $R(k,\infty)=k$ 
which is the expected large-$N$ limit.
A general  expression satisfying these conditions can be written as
\begin{equation}
G(k,N) = {U(k,N)\over U(1,N)},
\end{equation}
where
\begin{equation} 
U(k,N) =  \sin { k\pi\over N}  \;
\left[ 1 + \sum_{i=1} c_i(N)  \left( \sin {k\pi\over N} \right)^i \right]
\label{genf}
\end{equation}
and $c_i$ are coefficients which may depend on $N$, but in such a way
as not to spoil the large-$N$ limit, i.e. $c_i(N)/N^i=O(1/N)$.  Our
results show that, if the sine formula is not exactly satisfied, the
corrections must be small, thus $c_i\ll 1$.  In order to better
quantify this statement, one may keep only the first term in the sum
of Eq.~(\ref{genf}) and find a bound on $c_1$ (assuming it constant).
Our Monte Carlo results provide the bound $|c_1| \lesssim 0.05$. We
think that the accuracy of the sine formula in predicting the
$\sigma_k/\sigma$ ratio should trigger further fundamental theoretical
investigations.

\section{Strong-coupling expansion in the lattice Hamiltonian
approach of SU($N$) gauge theories}
\label{sec3}

The strong-coupling expansion in the lattice Hamiltonian formulation
of the theory is an analytical approach that can be used to
investigate Casimir scaling and its corrections.

The lattice Hamiltonian of a $D$-dimensional SU($N$) gauge theory is
defined in terms of link operators in the fundamental representation
$\hat{U}_\mu(x)$.  Following Ref.~\cite{KS-75}, we consider the
Hamiltonian
\begin{equation}
H_{\rm gauge} = {g^2\over 2a} \left\{ \sum_l \hat{E}^2(l) - 
{2\over g^4}  \sum_p {\rm Tr}
\left[ \hat{U}(p) + h.c. \right] \right\},
\label{Hgauge}
\end{equation}
where $\hat{E}^2(l)\equiv \sum_{a=1}^{N^2-1} \hat{E}^a(l)
\hat{E}^a(l)$ is the quadratic Casimir operator of SU($N$) associated
with the link $l$ of a $(D-1)$-dimensional cubic lattice, and
$\hat{U}(p)$ is the product of link operators on the boundary of a
plaquette $p$.  The SU($N$) gauge theory is realized in the continuum
limit $g\rightarrow 0$.  The lattice Hamiltonian formulation is not
unique, and therefore the corresponding strong-coupling expansion in
powers of $1/g$ should be considered as a suggestive investigation
method, but it is difficult to extract reliable quantitative
information from it.

In the perturbative strong-coupling approach in which $g\rightarrow
\infty$, one works in the space of states $\prod_x | U\rangle$
diagonal in $\hat{U}$:
\begin{eqnarray}
&&\hat{U}^{ab}_r | U\rangle = U^{ab}_r | U\rangle,\\
&&[\hat{E}^a,\hat{U}_r] = - {1\over 2}\lambda_r^a \hat{U},
\end{eqnarray} 
where $r$ indicates the generic representation and $\lambda_r^a$ the
corresponding generators.  In the strong-coupling limit $g\rightarrow
\infty$, the vacuum $|0\rangle $ is the lowest eigenstate of
\begin{equation}
H_0 \equiv  A \sum_l \hat{E}^2(l),
\end{equation}
where $A=g^2/2a$,
thus it satisfies $\hat{E}^2 |0\rangle = 0$.  In order to compute the
$k$-string tensions, we must consider the force law between widely
separated quarks in the strongly coupled limit. The potential energy
is defined as the lowest energy compatible with the presence of a
quark $q$ at the origin and an antiquark $\bar{q}$ at site $s$. The
minimum-energy gauge-invariant state is obtained by exciting the
shortest path of links connecting the $q\bar{q}$ pair.  Let us take
the $q\bar{q}$ pair along a side of the cubic lattice, thus the least
number of excited links is $l=s$.  In the strong-coupling limit
$g\rightarrow\infty$ it can be written as the product of $l$ link
operators joining the origin to the site $s$:
\begin{equation}
|r,l\rangle  = d_r^{-1/2}  \hat{U}_r(1) \hat{U}_r(2) ... \hat{U}_r(s)
 |0\rangle ,
\label{state}
\end{equation}
where $r$ is the representation of SU($N$) considered and $d_r$ its
dimension. The inner product $\langle | \rangle$ is defined using the
invariant integration over the group:
\begin{equation}
\int dg U_r^{ab}
\bar{U}_{r'}^{a'b'} = d_r^{-1} \delta_{r r'} \delta_{a a'}
\delta_{bb'}.
\end{equation}
The energy corresponding to the state $|r,l\rangle$ is given by the
matrix element \\ $\langle r,l| H_0 | r,l\rangle$. As already mentioned,
the leading order of the energy is proportional to $C_{r} =
\frac{1}{4}\lambda_r^a\lambda_r^a$, the value of the quadratic Casimir
operator in the representation $r$, i.e.
\begin{equation}
\langle r, l | H_0 | r, l\rangle = A \, l \, C_r .
\end{equation}
Higher order corrections can be computed by systematic application of
perturbation theory, where the perturbation interaction $H_I$ is given
by the second term of the Hamiltonian (\ref{Hgauge}).  The
perturbation $H_I$ corrects the energy of these states and of the
vacuum to $O(1/g^8)$, i.e. to second order in perturbation theory.
These corrections are obtained by computing
\begin{equation}
\langle r, l | H_I {1\over E-H_0} H_I | r,l \rangle
\end{equation}
and subtracting the corresponding second order contribution to the
vacuum energy
\begin{equation}
\langle 0 | H_I {1\over  -H_0} H_I | 0 \rangle .
\end{equation}
For the $k$-string tension associated with the rank-$k$ antisymmetric
representation $(1^k;0)$ (where $1^k$ indicates one column with $k$
squares in the corresponding Young tableau) we obtain
\begin{equation}
\sigma_k = A C_{(1^k;0)} \left[ 1 + 
 {D-2 \over (g^2 N)^4} e_4(k,N)  + O\left( {1\over (g^2 N)^6}\right)\right],
\label{sksc}
\end{equation}
where $C_{(1^k;0)}$ is the value of the quadratic Casimir operator for
the representation $(1^k;0)$, and
\begin{equation}
e_4(k,N) = {8N^4\over C_{(1^k;0)}}\left[
{1\over 2 C_{(1;0)}}   + 
\sum_{i=1}^4
{1 \over C_{(1^k;0)}  - 3 C_{(1;0)} - C_{r_i} } {d_{r_i}\over N d_{(1^k;0)}} \right].
\label{e4}
\end{equation}
The sum extends over the four representations $r_i$ obtained by
composing the rank-$k$ antisymmetric representation $(1^k;0)$ with the
fundamental $(1;0)$ and antifundamental $(0;1)$ representations,
i.e. $(1^{k+1};0)$, $(2,1^{k-1};0)$ and $(1^{k-1};0)$, $(1^{k};1)$
respectively. For notation see e.g. Ref.~\cite{PCV-98}.  The
corresponding dimensions $d_{r_i}$ and Casimir values $C_{r_i}$ are
\begin{eqnarray}
&&d_{(1^k;0)} = \left( \begin{array}{c} N\\ k\end{array}\right), \\
&&C_{(1^k;0)} = {k(N+1)(N-k)\over 2N}, \\
&&d_{(2,1^{k-1};0)} = k \, \left( \begin{array}{c} N+1\\ k+1\end{array}\right), \\
&&C_{(2,1^{k-1};0)} = {(k+1)[ N^2-N(k-2)- k-1]\over 2N}\, , \\
&&d_{(1^k;1)} = d_{(2,1^{N-k-1};0)} , \\
&&C_{(1^k;1)} = C_{(2,1^{N-k-1};0)}.
\end{eqnarray}
Notice that in the above expressions $(1^0;0)\equiv (0;0)$, i.e.  the
singlet representation.  For $k=1$, i.e. the fundamental string
tension, the expression (\ref{sksc}) reproduces the result reported in
Ref.~\cite{KPS-79}, i.e.
\begin{equation}
\sigma = A {N^2-1\over 2N} \left[ 1 - 
{D-2\over (g^2 N)^4} {16 N^6 (3N^2 -  5) \over (N^2-1)^2 (2N^2 - 1)(4N^2-9)}
+ ...\right].
\label{sfond}
\end{equation}
One can easily see that $e_4(k,N)$ satisfies the relation
$e_4(k,N)=e_4(N-k,N)$. It has the expected limit for $N\rightarrow
\infty$, indeed the term coming from the subtraction of the vacuum
contribution, which is the first one in Eq.~(\ref{e4}), cancels the
leading power in $N$, and the expression goes to a constant for
$N\rightarrow \infty$.

For the $k$-string tension ratio we obtain
\begin{equation}
{\sigma_k\over \sigma} = 
{k(N-k)\over N-1} \left\{ 1 + {D-2\over (g^2 N)^4} \left[ e_4(k,N) - e_4(1,N) \right] 
+ O\left( {1\over (g^2 N)^6}\right)\right\}.
\end{equation}
Explicitly for $k=2$,
\begin{eqnarray}
e_4(2,N) &-& e_4(1,N) = \nonumber \\
&&  \frac { 4N^3 (N-3) (12N^6 + 14N^5 - 46N^4 - 47N^3 +
61N^2 + 34N - 24)}{(N-2)(2N-3) (2N+3) (N^2-1)^2 (2N^2-1)(2N^2 +N
-4)} \nonumber \\
&=& {6\over N} - {2\over N^2} + O\left({1\over N^3}\right).
\label{k2k1}
\end{eqnarray}
Note that the correction to the leading order Casimir scaling is
$O(1/N)$.

This result provides an explicit example of corrections to the Casimir
scaling prediction: while the leading order strong coupling
calculation yields Casiir scaling, $O(1/N)$ corrections arise at the
next-to-leading order.  This fact holds for both $D=4$ and $D=3$.

\section{Chiral models and lattice gauge theories}
\label{sec2}

In this section we discuss the analogies existing between
$2d$-dimensional SU($N$) gauge theories and $d$-dimensional
SU($N$)$\times$SU($N$) chiral models.  In particular, we compare the
spectrum of the $k$-strings in SU($N$) gauge theories with the known
spectrum of the bound states in chiral models.

\subsection{Analogies}
\label{sec21}

Unitary matrix models defined on a lattice can be divided into two
major groups, according to the geometric and algebraic properties of
the dynamical variables: when the fields are defined in association
with lattice sites, and the symmetry group is global, i.e., a single
${\rm SU}(N)_L \times {\rm SU}(N)_R$ transformation is applied to all
fields, we are considering a spin model (principal chiral model); in
turn, when the dynamical variables are defined on the links of the
lattice and the symmetry is local, we are dealing with a gauge model
(lattice gauge theory).  An analogy between $d$-dimensional chiral
models and $2d$-dimensional gauge theories can be found according to
the following correspondence table \cite{GS-81,Polyakov-88}:
\begin{eqnarray*}
\begin{tabular}{c@{\quad}c}
{\bf spin} & {\bf gauge} \\
site, link & link, plaquette \\
loop & surface \\
length & area \\
mass $M$ & string tension $\sigma$ \\
two-point correlation & Wilson loop \\
\end{tabular}
\end{eqnarray*}
To this correspondence table, one may also add: the spectrum of the
bound states on the side of the chiral models and the spectrum of
confining strings for SU($N$) gauge theories, i.e.
\begin{eqnarray*}
\begin{tabular}{c@{\quad}c}
{\bf spin} & {\bf gauge} \\
bound state mass $M_k$ & $k$-string tension $\sigma_k$ \\
\end{tabular}
\end{eqnarray*}
While the above correspondence in arbitrary dimensions is by no means
rigorous, there is some evidence supporting the analogy.

Let us consider the following lattice formulation of $d$-dimensional
SU($N$)$\times$SU($N$) principal chiral models
\begin{equation}
S_{\rm chiral}=-2N\beta\sum_{x,\mu} 
{\rm Re}{\rm Tr}\left[ U(x) U^\dagger(x+\mu)\right]
\label{lchiralaction}
\end{equation}
(where $\beta={1/NT}$, $U_x\in {\rm SU}(N)$ and $\mu=1,...,d$); for
SU($N$) gauge theories in $2d$ dimensions we consider the Wilson
lattice formulation (\ref{wilsonac}).

In the case $d=1$ one can prove an identity between the partition
function (and appropriate correlation functions) of the
two-dimensional lattice gauge theory and the corresponding quantities
of the one-dimensional principal chiral model (see
e.g. Ref.~\cite{PCV-98}).  Both theories are exactly solvable and the
correspondence can be explicitly shown.  In particular, using the
results of Ref.~\cite{Rossi-81}, one may easily show that the Casimir
formula holds for the spectrum. Thus
\begin{equation}
{M_k\over M} = {k(N-k)\over N-1} 
\end{equation}
holds for the one-dimensional SU($N$)$\times$SU($N$) chiral models,
and
\begin{equation}
{\sigma_k\over \sigma} = {k(N-k)\over N-1} 
\end{equation}
for two-dimensional SU($N$) gauge theories.

Although for higher dimensions the correspondence does not strictly
hold, we still have some analogies:

(i) The two-dimensional chiral model and the four-dimensional
non-abelian gauge theory share the property of asymptotic freedom and
dynamical generation of a mass scale.  In both models these properties
are absent in the Abelian case (${\rm XY}$ model and ${\rm U}(1)$
gauge theory respectively).

(ii) The classical equations of motion describing the dynamics of the
spin variables in the two-dimensional chiral model and the Wilson
loops in the four-di\-men\-sio\-nal nonabelian gauge theory are similar (see
e.g. Ref.~\cite{Polyakov-88}).  It turns out that the gauge fields are
chiral fields on the loop space. This analogy persists at the quantum
level as well, since the classical equations of motion dictate
relations among correlation functions in the corresponding quantum
theory.

(iii) Exploiting asymptotic freedom, one can use perturbation theory
to determine the short-distance behavior of the potential
of two heavy quarks in four-dimensional SU($N$) gauge theories and
of the two-point correlation function in two-dimensional
SU($N$)$\times$SU($N$) chiral models. Both show Casimir scaling.
In the four-dimensional SU($N$) gauge theory
the force $F_r(x)$ between two heavy
quarks in the representation $r$ is proportional to $C_r$ up to two
loops \cite{pert-pot}.  Resumming the leading logarithms using
standard renormalization-group arguments, the short-distance behavior
of the force $F_r(x)$ is:
\begin{eqnarray}
&&F_r(x) = - C_r {\alpha_r(1/x)\over x^2},\nonumber \\ &&
\alpha_r(1/x) = {1\over b_0 L(x)} \left[ 1 - b_1 {\ln L(x)\over L(x)}
+ O\left( {(\ln L)^2\over L^2}\right)\right],\nonumber
\end{eqnarray}
where $L(x) = - \ln x \Lambda_x$ and $\Lambda_x$ is a mass scale (the
so-called $\Lambda$-parameter associated with the above definition of
running couplings $\alpha_r(x)$; $b_0$ and $b_1$ are the first two
universal coefficients of the perturbative expansion of the
$\beta$-function $\beta(\alpha) = - x \partial \alpha_r/\partial x$.
$b_0$, $b_1$ and $\Lambda_x$ are independent of $C_r$.  A dependence
on $C_r$ of the $O\left( (\ln L)^2/L^2\right)$ term in $\alpha_r(1/x)$
is not excluded. Its computation requires a three-loop calculation of
the potential.  As we shall see in Sec.~\ref{sec22}, Casimir
scaling emerges also in the short-distance behavior of two-point
functions in chiral models.

(iv) Another analogy concerns the large-$N$ limit, which in both cases
is given by a sum of planar graphs. In this limit particles in chiral
theories become free, indeed the known S-matrix
\cite{AAL-84,Wiegmann-84} becomes trivial. On the other hand, it is
conjectured that the large-$N$ limit of gauge theories is a free
string theory, see e.g. \cite{Polyakov-88}, although its nature is
still not clear.

(v) Their strong-coupling expansions look similar, provided that
point-like excitations of chiral fields are substituted by closed flux
lines for the gauge fields.

(vi) Approximate real-space renormalization recursion relations
obtained by Mig\-dal \cite{Migdal-75} are identical for $d$-dimensional
chiral models and $2d$-dimensional gauge models.

\subsection{The two-dimensional SU($N$)$\times$SU($N$) chiral models}
\label{sec22}

The two-dimensional SU($N$)$\times$ SU($N$) principal chiral models
are asymptotically free matrix-valued field theories defined by the action
\begin{equation}
S={1\over T} \int d^2x {\rm Tr} \;\partial_\mu U(x) \, \partial_\mu U^\dagger(x).
\label{caction}
\end{equation}
where $U(x)\in$ SU($N$).

Using the existence of an infinite number of conservation laws and
Bethe-Ansatz methods, the on-shell solution of the ${\rm SU}(N)\times
{\rm SU}(N)$ chiral models has been proposed in terms of a factorized
S-matrix \cite{AAL-84,Wiegmann-84}.  The analysis of the corresponding
bound states leads to the mass spectrum
\begin{equation}
M_k=M{\sin(k\pi/N) \over \sin(\pi/N)},
\qquad 1\leq k\leq N-1,
\label{masses}
\end{equation}
where $M_k$ is the mass of the $k$-particle bound state transforming
as a totally antisymmetric tensor of rank $k$.  $M\equiv M_1$ is the
mass of the fundamental state determining the Euclidean long-distance
exponential behavior of the two-point Green's function
\begin{equation}
G(x)= {1\over N}\langle {\rm Tr} \,U(0) U(x)^\dagger \rangle.
\label{fgf}
\end{equation} 
No bound states exist for other representations. Correlation
functions associated with the generic representation $r$ can be
defined by
\begin{equation}
G_r(x)= {1\over d_r}\langle \chi_r\left[ U(0) U(x)^\dagger \right] \rangle,
\label{grx}
\end{equation}
where $d_r$ and $\chi_r$ are respectively the dimension and the
character of the representation $r$.  The structure of the S-matrix
implies that stable bound states propagate only in the totally
antisymmetric representations with masses given by Eq.~(\ref{masses}).
Note that, according to the correspondence discussed in
Sec.~\ref{sec21}, the correlation functions $G_r(x)$ play the role of
the $r$-representation Wilson loops for SU($N$) gauge theories.  The
mass spectrum (\ref{masses}) has been verified numerically at $N=6$ by
Monte Carlo simulations \cite{RV-94,DH-94}.

As in SU($N$) gauge theories, the large-$N$ limit of these models is
formally represented by a sum over planar graphs.  The $S$-matrix has a
convergent expansion in powers of $1/N$, and becomes trivial, i.e.,
the $S$-matrix of free particles, in the large-$N$ limit.  Note also
that, as in the case of SU($N$) gauge theories, no a priori reasons
exist for the large-$N$ expansion of the $k$-state mass ratios
(\ref{masses}) to be even in $1/N$.

Asymptotic freedom allows us to determine the small-distance behavior
of the correlation functions in perturbation theory. In
two-dimensional SU($N$)$\times$SU($N$) chiral models, the logarithm of
the two-point function $G_r(x)$ is the analog of the potential of two
separated quarks in the representation $r$.  The small-distance
behavior of $\ln G_r(x)$ satisfies Casimir scaling, similarly to what
happens for the potentials in four-dimensional SU($N$) gauge theories.
Extending the results of Ref.~\cite{MS-80} pertaining to the
fundamental two-point function, in the $\overline{MS}$ renormalization
scheme associated with the $x$ space we find
\begin{eqnarray}
&&\ln G_r(x) =  - C_r t \Big\{  \ln(\mu x) + {N\over 8}  t \ln(\mu
x)^2 
\label{lngr} \\
&&+t^2 \Big[   {N^2\over 256} \Big(3 -2\zeta(3)\Big)  + {3N^2\over 128} \ln(\mu x)
+{N^2\over 64} \ln(\mu x)^2 + {N^2\over 48} \ln(\mu x)^3 \Big] + O(t^3)\Big\},
\nonumber
\end{eqnarray}
where $C_r$ is the Casimir value of the representation $r$. As in
four-dimensional SU($N$) gauge theories, one may define a running
coupling $t_x$ from the relation
\begin{equation}
-{\partial \ln G_r(x)\over \partial x} = C_r {t_x\over x}.
\end{equation}
Standard renormalization-group arguments allow one to resum the leading
logarithms, yielding
\begin{equation}
t_x = {1\over b_0 L(x) }\; 
\left[ 1 - {b_1 \ln L(x)\over L(x)}  + O\left({(\ln L)^2 \over L^2} \right) \right].
\label{tx}
\end{equation}
As before, $L(x) = -\ln x\Lambda_x$ and $\Lambda_x$ is a mass scale;
$b_0=N/(8\pi)$, $b_1=N^2/(128\pi^2)$ are the first two coefficients of
the $\beta$-function $\beta(t_x) = -x{\partial t_x /\partial x} = -
b_0 t_x^2 - b_1t_x^3 + O(t_x^4)$.  We also mention that at the next
order of perturbation theory, i.e. $O(t^4)$ in Eq.~(\ref{lngr}), there
are three-loop diagrams whose group factor would violate Casimir
scaling.  Thus, barring particular cancellations, Casimir scaling
should not be satisfied by the $O\left({(\ln L)^2/L^3} \right)$ term
of Eq.~(\ref{tx}).  The same observation applies to four-dimensional
SU($N$) gauge theories.

Finally, again similarly to QCD, one may easily check that, using the
lattice Hamiltonian approach of Ref.~\cite{SK-81}, Casimir scaling is
recovered in the strong-coupling limit; it is violated by higher order
corrections.

The analogy with chiral models highlights some general trends: Casimir
scaling is exact in $d=1$, as it is in the case of SU($N$) Yang Mills
theories in $d=2$; for $d=2$, both perturbation
theory and strong coupling yield Casimir scaling to lowest order. However
next-to-leading order calculations explicitly show corrections to
such behavior. This fact is consistent with the picture that emerges
for $d=4$ SU($N$) gauge theories from our Monte Carlo simulations.

\acknowledgments

\noindent
We thank M. Campostrini, K. Konishi, S. Lelli, B. Lucini, M. Maggiore,
A. Pelissetto, V. Shevchenko, and M.  Teper for useful and interesting
discussions.  LDD enjoyed the warm hospitality of the Theoretical
Physics Department in Oxford, where some of the above discussions took
place.  H. P. would like to thank the Theory Group in Pisa for their
hospitality during various stages of this work.
We thank Maurizio Davini for setting up the cluster that ran the
simulations and for his indispensable technical support.

\appendix

\section{Rescaled and effective lattice couplings}
\label{appa}

In order to compare Monte Carlo results for various values of $N$, it
is useful to introduce the rescaled coupling
\begin{equation}
\gamma \equiv {1\over g_0^2 N} = {\beta \over 2 N^2}
\end{equation}
which is kept fixed in the large-$N$ limit of the lattice theory, see
e.g. Refs. \cite{Das-87,PCV-98}.  As already noted 
in Refs.~\cite{Teper-97,LT-01-2}, the correspondence of the bare
couplings for models with different values of $N$ (defined keeping
physical quantities such as the string tension fixed) becomes more
accurate if one uses the mean field improved coupling $g_{\rm mf}$
proposed in Refs.~\cite{improved-couplings,LM-93} and obtained by
dividing the lattice coupling $g_0^2$ by the average value of the
plaquette, i.e.
\begin{equation}
g_{\rm mf}^2 = {g_0^2\over \langle \frac{1}{N} {\rm Tr}\,
U_{\mu\nu}(x)\rangle}.
\end{equation}
Here we also consider the effective coupling $g^2_{\rm cactus}$
obtained within an all-order resummation of cactus-type diagrams in
perturbation theory \cite{PV-98}, and defined by
\begin{equation}
g^2_{\rm cactus} = {g_0^2\over w(g_0)}.
\end{equation}
The function $w(g_0)$ can be extracted by an appropriate
algebraic equation that can be easily solved numerically:
\begin{eqnarray}
&&u \, e^{-u (N{-}1)/(2N)} \,\left[ {N{-}1\over N} \, L^1_{N{-}1}(u) +2
L^2_{N{-}2}(u) \right] = {g_0^2\,(N^2{-}1)\over 4}, \label{zGeq}\\
&&u(g_0) \equiv {g_0^2 \over 4 (1{-}w(g_0))}\, , \nonumber
\end{eqnarray}
where $L^M_N$ are the Laguerre polynomials.  Data for $N=3,4,6$ are
shown in Fig.~\ref{effcou} (the data for $N=3$ and some of those for
$N=4$ have been taken from Ref.~\cite{LT-01-2}): we plot $1/\gamma=N
g_0^2$, $1/\gamma_{\rm mf}=N g^2_{\rm mf}$ and $1/\gamma_{\rm cactus}=
N g^2_{\rm cactus}$ versus $a\sigma^{1/2}$, where $\sigma$ is the
string tension.  The gray points, corresponding to the mean field
improved coupling, fall approximately on a single curve. A similar
behaviour is observed for the cactus improved coupling (black points),
while the data corresponding to the bare coupling (white points) show
a wider spread as $N$ is varied. We conclude that both the improved
couplings provide an efficient tool to match theories at different
values of $N$.  In this respect, the mean field improved coupling
performs slightly better, but the advantage of the cactus definition
is that it is determined from Eq.~(\ref{zGeq}), without requiring any
simulation.

\FIGURE[ht]{
\epsfig{file=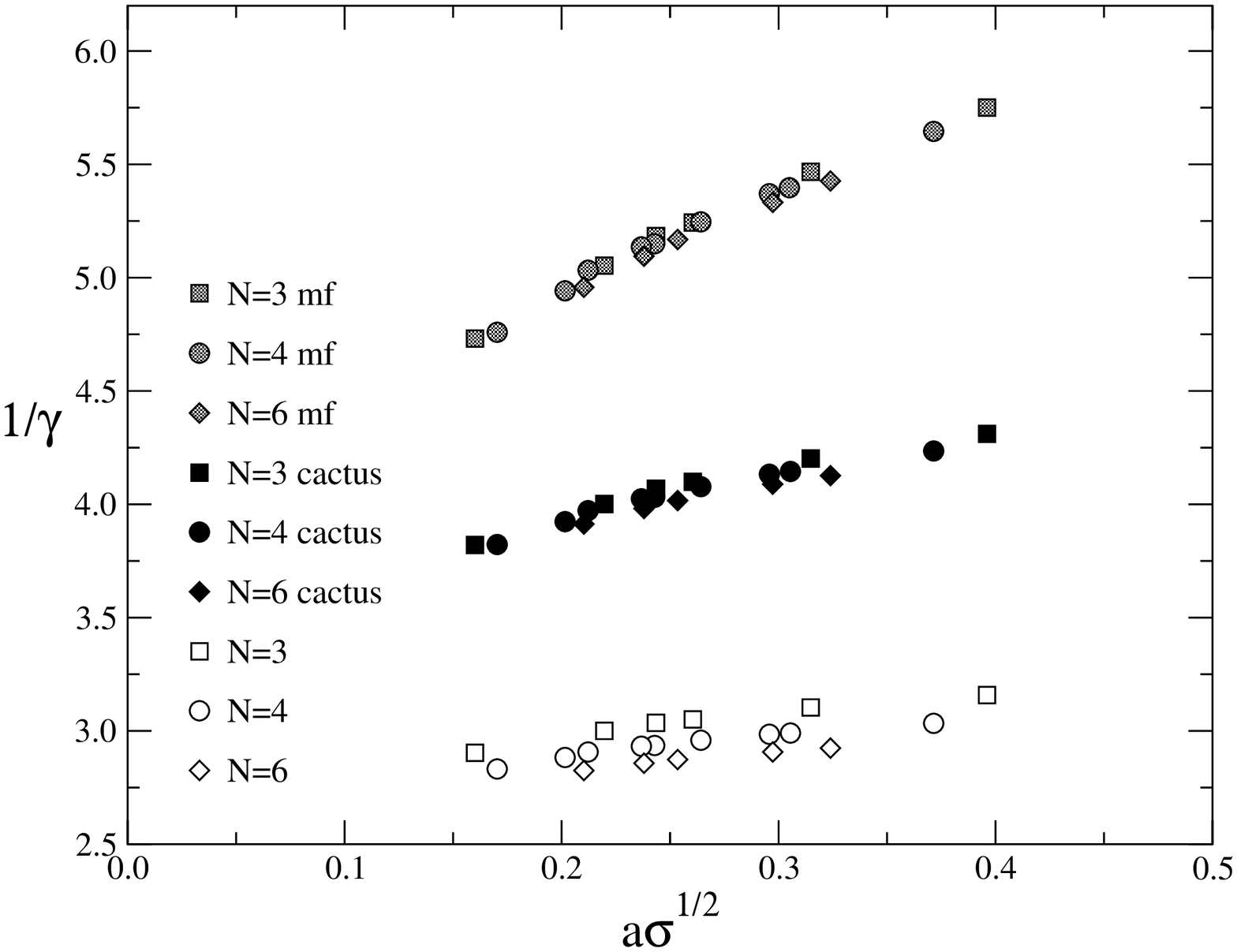, width=12truecm}
\caption{$1/\gamma$, $1/\gamma_{\rm mf}$ and $1/\gamma_{\rm cactus}$
versus $a\sigma^{1/2}$ for SU(3), SU(4), and SU(6) lattice gauge
theories.}
\label{effcou}
}

\section{Bulk phase transitions at large $N$}
\label{appb}

For sufficiently large values of $N$ and in particular in the
large-$N$ limit, the Wilson lattice formulation of SU($N$) gauge
theories presents a first order phase transition.  This has been
argued using various approaches, such as Monte Carlo simulations
\cite{Creutz-81,CrMo-82,others-PT,IIY-95}, mean field calculations
\cite{DZ-83,ID-89}, reduced models \cite{Campostrini-99}.  In the
following we present evidence for a first order transition at a finite
bare coupling in the case of the SU(6) lattice gauge theory.  On the
other hand, in the SU(4) case no evidence of a bulk phase transition
is found.

\subsection{A first order transition for the SU(6) lattice gauge theory}

In the case $N=6$, we performed simulations
starting from hot and cold configurations.  Fig.~\ref{hysteresis}
reports data for the energy density $E$ (normalized to one for
$\gamma=0$) obtained performing two simulations on a $8^4$ lattice,
one starting from a hot configuration and the other from a cold
configuration.  More precisely, in the first case we started from a
hot configuration, and performed simulations starting from
$\gamma=0.334$, increasing the value of $\gamma$ every 2000 heat bath
updatings by 0.001.  The data reported in the figure are the values of
the energy obtained averaging over the second 1000 updatings for each
$\gamma$ value. The data concerning the simulation starting from a
cold configuration were obtained similarly, i.e. starting from
$\gamma=0.344$ and decreasing its value by 0.001 every 2000
upgradings.  Fig.~\ref{hysteresis} shows clearly the presence of hysteresis,
from which one may estimate $\gamma_c\approx 0.339$.  Note
that the latent heat is relatively large indicating that the first
order transition is rather strong.  Another estimate of $\gamma_c$ can
be obtained from the so-called mixed-phase method. One starts from a
configuration that is half cold and half hot, and see which phase wins
as $\gamma$ is varied.  An estimate of $\gamma_c$ is obtained from the
boundary values of $\gamma$ for which the change of final phase
occurs.  The two phases are easily recognized by their value of the
energy: as shown in Fig.~\ref{hysteresis}, $E_{\rm hot}\simeq 0.57$
and $E_{\rm cold}\simeq 0.47$ at $\gamma_c$. We obtained
$\gamma_c\approx 0.3393$, $\gamma_c\approx 0.3390$ and
$\gamma_c\approx 0.3389$ respectively from simulations on $8^4$,
$10^4$ and $12^4$ lattices.  As a final estimate we consider
$\gamma_c=0.3389(4)$.  For comparison, we mention the earlier estimate
$\gamma_c=0.333(14)$ \cite{CrMo-82}.

\FIGURE[ht]{
\epsfig{file=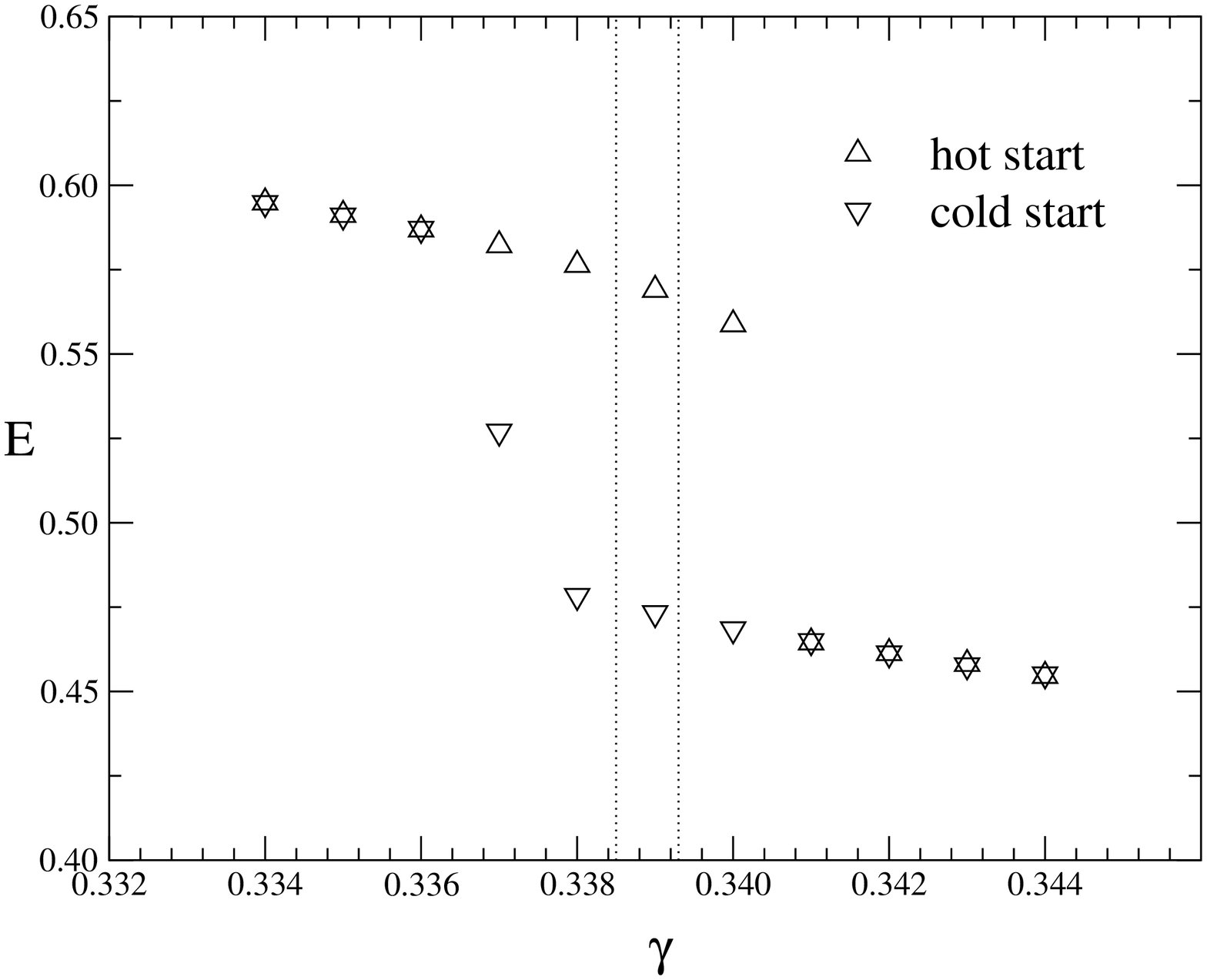, width=12truecm}
\caption{Energy density versus $\gamma$ obtained as explained in the
text. The two vertical lines show the estimate of $\gamma_c$ obtained
by the mixed-phase method.}
\label{hysteresis}
}

\subsection{Just a crossover  for the SU(4) lattice gauge theory}

In the case of the SU(4) lattice gauge theory, no evidence for a bulk
transition is observed.  In particular, the specific heat does not
appear to diverge with increasing lattice size.  In
Fig.~\ref{Ch4} we plot data for the specific heat
\begin{equation}
C_H = - \beta^2 {d\over d\beta} \left( 1 - {1\over N}
\langle {\rm Tr}\; U_{\mu\nu}(x) \rangle \right)
\end{equation}
for various lattice sizes, i.e. $6^4$, $8^4$ and $12^4$.  The data
were obtained from runs with typical statistics of 3-5$\times 10^4$
sweeps.  They show a rather pronounced peak around $\gamma\simeq
0.325$ (corresponding to $\beta=10.4$), but they appear to converge
for increasing lattice size. We recall that, in the case of a first
order phase transition, the finite-size scaling behavior of the peak
value of the specific heat should diverge as \cite{CLB-86}
\begin{equation}
C_{H,{\rm peak}} \sim L^d, 
\end{equation}
thus $C_{H,{\rm peak}} \sim L^4$ in our case.  The data of
Fig.~\ref{Ch4} are clearly inconsistent with such a behavior.  These
results indicate that the Wilson formulation of the SU(4) lattice
gauge theory does not undergo a first-order phase transition, but
rather they suggest that, as in the SU(3) case, it exhibits a
crossover between the strong and weak coupling regime, characterized
by a rather pronounced peak of the specific heat.

We mention that early works, based on mean field calculations
\cite{ID-89} and Monte Carlo simulations \cite{Creutz-81,others-PT},
suggested a bulk first-order phase transition also for SU(4) and
$\beta\simeq 10.4$.

\FIGURE[ht]{
\epsfig{file=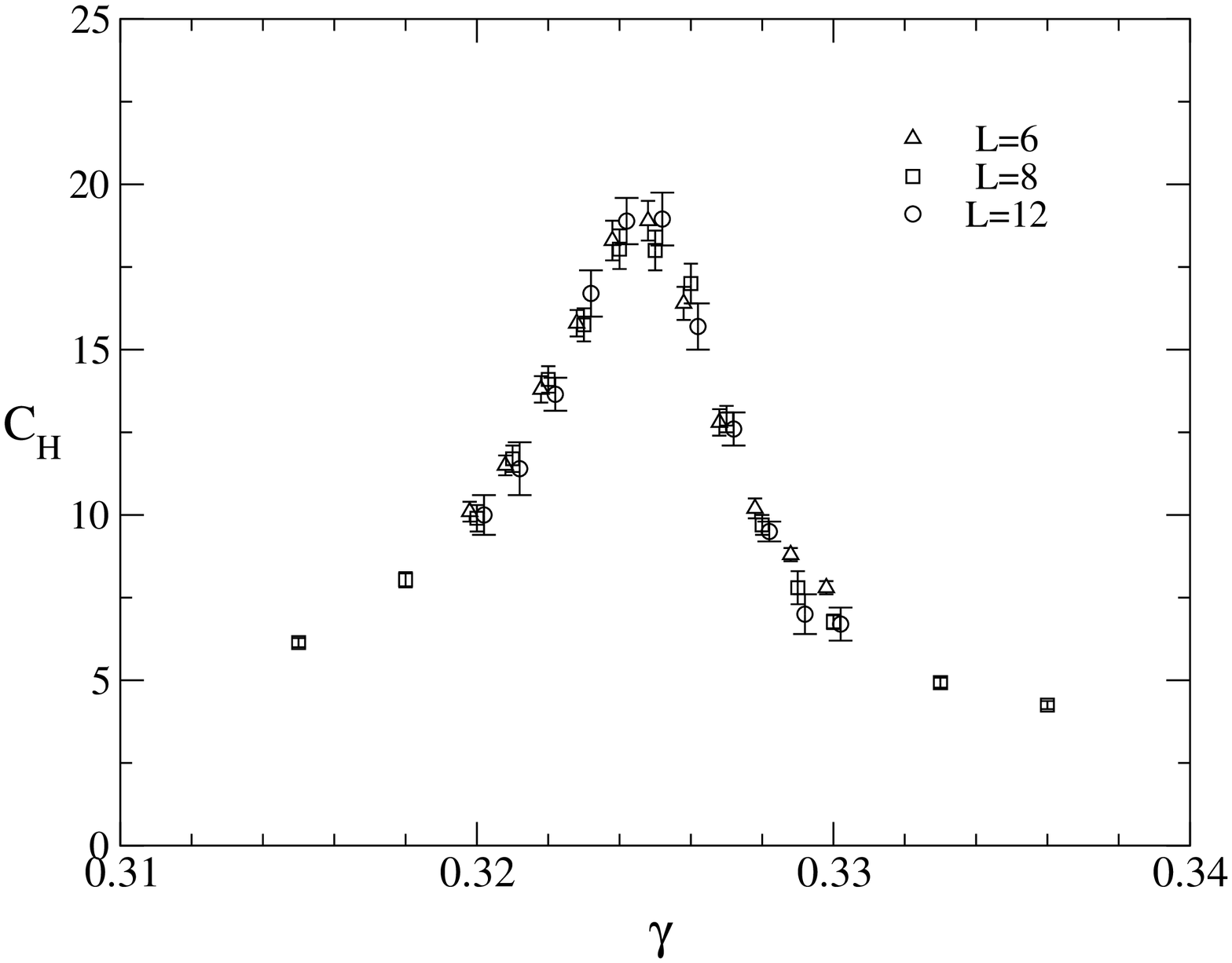, width=12truecm}
\caption{Specific heat versus $\gamma$ for $SU(4)$ as obtained from
simulations on $6^4$, $8^4$ and $12^4$ lattices.  Data for $L=6,12$
are slightly shifted horizontally to make the
figure more readable.}
\label{Ch4}
}

\section{Critical slowing down for the topological modes}
\label{appc}

Monte Carlo simulations of critical phenomena in statistical mechanics
and of the continuum limit in quantum field theory are hampered by the
problem of critical slowing down.  For a general introduction to
critical slowing down in Monte Carlo simulations, see
e.g. Ref.~\cite{Sokal-92}.  The autocorrelation time $\tau$, which
corresponds to the number of iterations needed to generate a new
independent configuration, grows with increasing length scale $\xi$;
usually it blows up following a power law, i.e. $\tau\sim \xi^z$. In
gauge theories one may consider $\xi_{\sigma}\equiv \sigma^{-1/2}$ as
a length scale.  Critical slowing down for traditional algorithms,
such as standard Metropolis or heat bath, arises essentially from the
fact that their updating is local: in a single step of the algorithm,
information is transmitted from a given site/link to the nearest
neighbors.  Roughly, one expects that this information spreads
following a random walk around the lattice.  An essentially
independent configuration is obtained when the information travels a
distance of the order of the correlation length $\xi$. This suggests
that $\tau\sim\xi^2$. This guess is correct for Gaussian (free field)
models; in general one expects that $\tau\sim \xi^z$ where $z$ is a
dynamical critical exponent.  In the case of local algorithms, such as
Metropolis and heat bath, one expects $z\simeq 2$.  Appropriate
overrelaxation procedures may achieve a reduction of $z$, although the
condition $z\ge 1$ holds for local algorithms.  On the other hand, in
the presence of relevant topological modes, the random-walk picture
may fail.  These modes may give rise to sizeable free-energy barriers
separating different regions of the configuration space.  The
evolution in the configuration space may present a long-time
relaxation due to transitions between different topological charge
sectors.  (Although lattice field theories cannot strictly possess
topological properties, these are expected to be recovered in the
continuum limit.)  As in glass models, see e.g. Ref.~\cite{Parisi-92},
and in liquids below a crossover transition, see
e.g. Ref.~\cite{SSDG-00}, the presence of significant free-energy
barriers may determine an effective separation of short-time
relaxation within the free-energy basins from long-time relaxation
related to the transitions between basins.  The mechanism underlying
this long-time relaxation is rather different from the simple
random-walk spread of information, so the autocorrelation time of the
topological modes may not show a simple power law behavior.  This
picture was also behind the so-called heating method \cite{DV-92} to
measure the lattice renormalizations of topological charge
operators. This method exploits the critical slowing down of the
physical topological modes in off-equilibrium simulations, to
disentangle them from the short-distance lattice renormalizations.

Let us consider an observable $O$; its autocorrelation function
$C_O(t)$ ($t$ is the Monte Carlo time, i.e. the integer counting
Monte Carlo iterations at equilibrium) is defined as
\begin{equation}
C_O(t) = \langle \left( O(t) - \langle O \rangle \right)
\left( O(0) - \langle O \rangle \right) \rangle,
\end{equation}
where the averages are taken at equilibrium.
The integrated autocorrelation time $\tau_O$ associated with $O$ is 
given by
\begin{equation}
\tau_O = {1\over 2} \sum_{t=-\infty}^{t=+\infty} {C_O(t)\over C_O(0)}.
\end{equation}
Estimates of $\tau_O$ can be obtained by a blocking analysis of the
data, without measuring the autocorrelation function $C_O(t)$.
Indeed the following relation holds
\begin{equation}
\tau_O = N_{\rm m} \,{E^2\over 2 E_0^2},
\label{eqbl}
\end{equation}
where $N_{\rm m}$ is the number of sweeps between two measurements
of the observable $O$, $E_0$ is the naive error calculated without
taking into account the autocorrelations, and $E$ is the correct error
found after the blocking procedure (the estimate is meaningful only
if $N_{\rm m}\lesssim \tau_O$).
Of course $\tau_O$ depends on the observable $O$; the largest
value $\tau_O$ among the various observables provides the
time scale to obtain a new independent configuration in 
simulations at equilibrium.
In Ref.~\cite{TARO-94} the autocorrelation time $\tau_w$
of small Wilson loops was found to behave 
approximately as $\tau_w \sim \xi^2$ for the SU(3) gauge theories,
see also Ref.~\cite{GW-99}.
As we shall see, a more severe form of critical slowing down 
is observed in measuring quantities related to the topological
modes, such as  the topological charge $Q$.

We used cooling
to determine  $Q$ from each lattice configuration, measuring $Q$
every $N_{\rm m}=100$ Monte Carlo sweeps.
Estimates of the (integrated) autocorrelation time $\tau_Q$ were
obtained by a blocking analysis of the data, using Eq.~(\ref{eqbl}).
In the SU(6) case we found
$\tau_Q \approx 268$ for $\gamma=0.342$, $\tau_Q \approx 466$ for
$\gamma=0.344$, $\tau_Q \approx 1750$ for $\gamma=0.348$, and $\tau_Q
\gtapprox 3000$ for $\gamma=0.350$.  Moreover we found that the run at
the largest value of $\gamma$ considered, i.e. $\gamma=0.354$, did not
correctly sample the topological charge 
presumably because the expected $\tau_Q$ is large and the run was not
sufficiently long (approximately 300k sweeps). In the  SU(4) case we found
$\tau_Q\approx  101$ for $\gamma=0.338$,
$\tau_Q\approx  210$ for $\gamma=0.341$, and
$\tau_Q\approx  410,434$ for $\gamma=0.344$ and $L=16,12$.
The uncertainty on the above numbers should be at most
5\% for SU(4) and 10-20\% for SU(6).

These estimates of $\tau_Q$
are suggestive of an interesting phenomenon. 
As shown in Fig.~\ref{CSD}, they are consistent
with an exponential critical slowing down.
Indeed the data for the autocorrelation time
can be well fitted by an exponential behavior of the type
$\tau_Q\propto \exp (c \xi_\sigma)$ with $c\approx 2.5$
for SU(6) and $c\approx 1.7$ for $SU(4)$. 
One might also guess an increase of the constant $c$
according to $c\sim N$. 
We also mention that the data for $\tau_Q$ are
definitely inconsistent with a behavior of the type $\xi_\sigma^z$ with
$z\simeq 2$. Indeed, an acceptable power law behavior
fitting reasonably well the data for $\tau_Q$ would
require $z\simeq 7$ for SU(4) and $z\simeq 9$ for SU(6).
We expect that a similar critical slowing down phenomenon
occurs also for SU(3), and more generally in the presence of dynamical fermions.
 
This dramatic effect was not observed 
for the other quantities considered in our 
study, such as the correlations determining 
the $k$-string tensions and the glueball masses.
A blocking analysis of the data 
did not show significant time correlations in measurements taken every
10-20 sweeps for all values of $\gamma$ considered.  
For instance, in the case of SU(4) and for $\gamma=0.344$ and  $L=16$
the autocorrelation time $\tau_P$  of  smeared and blocked Polyakov line correlators 
was estimated to be smaller than 10, more precicely $\tau_P\approx 6$,
by a blocking analysis of the data.
This fact suggests an approximate
decoupling between the topological and nontopological modes.  
This seems to be also supported by the fact that string tension
results for $\gamma=0.354$ in the SU(6) case (see Table~\ref{tab:results6}),
extracted from a simulation which did not sample correctly $Q$, 
turn out to be in agreement with those for smaller
$\gamma$, for which $Q$ was sampled correctly.

\FIGURE[ht]{
\epsfig{file=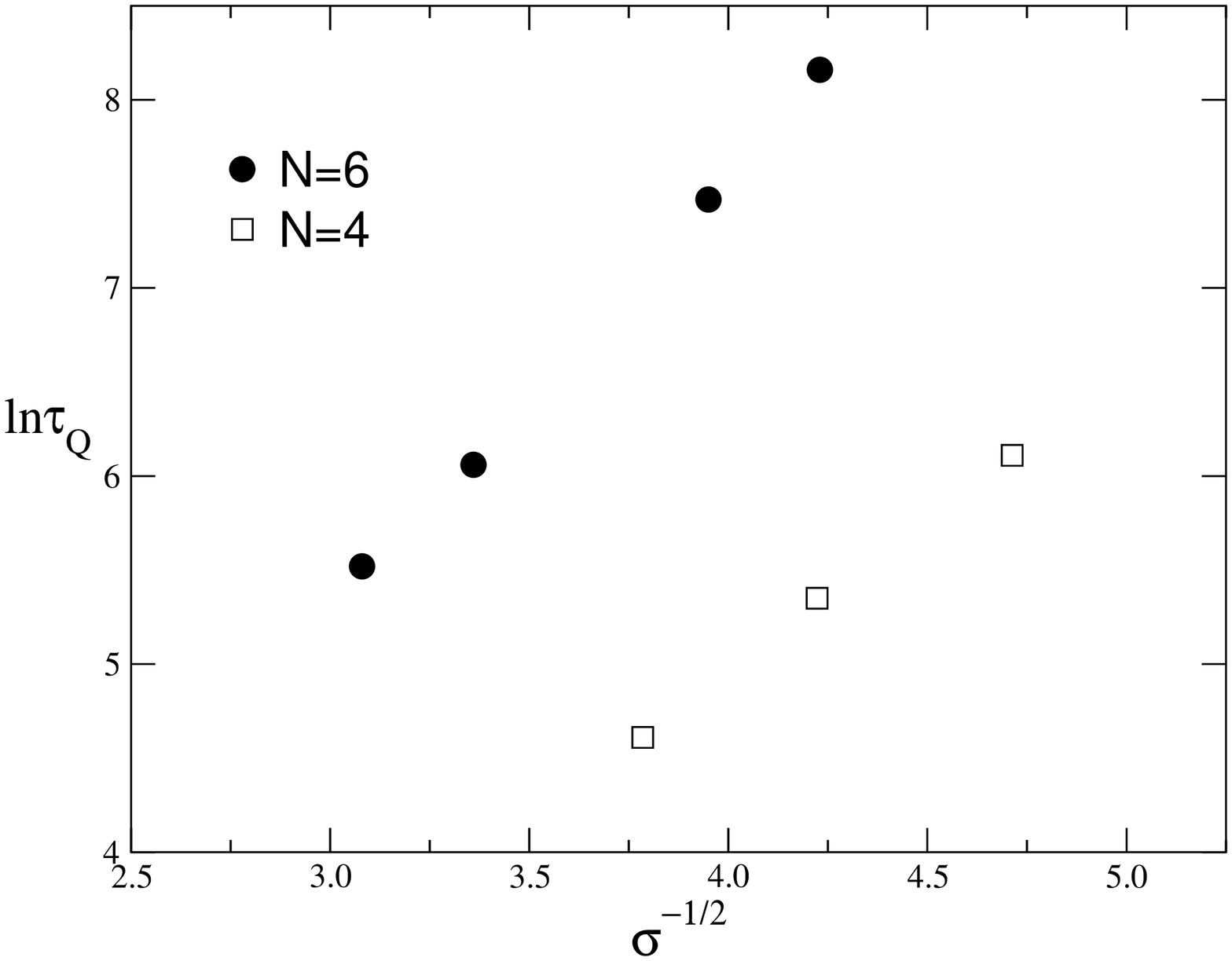, width=12truecm}
\caption{Plot of the autocorrelation time 
$\tau_Q$ of the topological charge versus $\sigma^{-1/2}$.
}
\label{CSD}
}

Such a critical slowing down phenomenon was already observed in simulations
of two-dimensional CP$^{N-1}$ models \cite{CRV-92,V-93}. Given that
these models possess some of the properties expected to hold
in QCD \cite{DDL-79,Witten-79} (asymptotic freedom, a form of
confinement due to the $U(1)$ gauge invariance, a nontrivial
topological structure), they have been used as a
laboratory to check and develop methods to investigate topological
properties.   Monte Carlo studies of 
lattice formulations of two-dimensional 
CP$^{N-1}$ models, using local
algorithms (mixtures of overrelaxation and heat bath upgradings,
similar to the ones generally used for the four-dimensional SU($N$) gauge theories), 
have shown that the critical slowing down of the topological modes 
is consistent with an exponential growing with respect to the
correlation length, worsening with increasing $N$.  On the other hand,
nontopological quantities, such as the mass gap and Wilson loops,
turned out not to be affected by this problem, suggesting a large
decoupling between the topological and nontopological modes.

\end{document}